\documentclass[twocolumn,trackchanges]{aastex7}

\usepackage{amsmath}
\usepackage{bm}

\newcommand{\arepo}{\texttt{AREPO}}
\newcommand{\mesa}{\texttt{MESA}}

\newcommand{\Msun}{\,\mathrm{M}_\odot}
\newcommand{\Rsun}{\,\mathrm{R}_\odot}
\newcommand{\Lsun}{\,\mathrm{L}_\odot}

\newcommand{\hr}{\,\mathrm{hr}}
\newcommand{\yr}{\,\mathrm{yr}}
\newcommand{\kyr}{\,\mathrm{kyr}}
\newcommand{\Myr}{\,\mathrm{Myr}}

\newcommand{\K}{\,\mathrm{K}}
\newcommand{\G}{\,\mathrm{G}}
\newcommand{\kG}{\,\mathrm{kG}}
\newcommand{\erg}{\,\mathrm{erg}}
\newcommand{\kmsinv}{\,\mathrm{km}\,\mathrm{s}^{-1}}
\newcommand{\gcmcbinv}{\,\mathrm{g}\,\mathrm{cm}^{3}\,\mathrm{s}^{-1}}

\newcommand{\ms}{m_{\star}}
\newcommand{\msone}{m_{\star 1}}
\newcommand{\mstwo}{m_{\star 2}}
\newcommand{\rs}{r_{\star}}
\newcommand{\rsone}{r_{\star 1}}
\newcommand{\rstwo}{r_{\star 2}}
\newcommand{\rtot}{r_{\mathrm{tot}}}
\newcommand{\Ls}{L_{\star}}
\newcommand{\Zs}{Z_{\star}}
\newcommand{\Xc}{X_{\mathrm{core}}}
\newcommand{\tdyn}{t_{\mathrm{dyn,\star}}}
\newcommand{\tth}{t_{\mathrm{th,\star}}}
\newcommand{\tms}{t_{\mathrm{MS,\star}}}
\newcommand{\tdynrtot}{t_{\mathrm{dyn},r_{\mathrm{tot}}}}
\newcommand{\Ncells}{N_{\mathrm{cells}}}
\newcommand{\rp}{r_{\mathrm{p}}}

\newcommand{\Bsurf}{B_{\mathrm{surf}}}

\newcommand{\Bphi}{B_{\phi}}
\newcommand{\Bravg}{\overline{B}_{r}}
\newcommand{\EB}{E_{\mathrm{B}}}
\newcommand{\Eturb}{E_{\mathrm{turb}}}
\newcommand{\mbound}{m_{\mathrm{\star,bound}}}
\newcommand{\Etot}{E_{\mathrm{\star,tot}}}
\newcommand{\Lax}{L_{\mathrm{\star,ax}}}
\newcommand{\vrad}{v_{\mathrm{rad}}}
\newcommand{\vradax}{v_{\mathrm{rad,ax}}}
\newcommand{\betamag}{\beta_{\mathrm{mag}}}
\newcommand{\omegamid}{\omega_{\mathrm{mid}}}
\newcommand{\omegacrit}{\omega_{\mathrm{crit}}}

\shorttitle{Stellar collisions and mergers}
\shortauthors{Vynatheya et al.}
\received{XX}
\revised{YY}
\accepted{ZZ}
\submitjournal{ApJ}

\begin{document}

\title{The collision and merger products of stars do not look alike: A magnetohydrodynamics comparison}

\author[0000-0003-3736-2059]{Pavan Vynatheya}
\affiliation{Canadian Institute for Theoretical Astrophysics, University of Toronto, 60 St George St, Toronto, ON M5S 3H8, Canada}
\affiliation{Max-Planck-Institut für Astrophysik, Karl-Schwarzschild-Straße 1, 85748 Garching bei München, Germany}
\email{pavanvyn@cita.utoronto.ca}

\author[0000-0003-2012-5217]{Taeho Ryu}
\affiliation{JILA, University of Colorado and National Institute of Standards and Technology, 440 UCB, Boulder, 80308 CO, USA}
\affiliation{Department of Astrophysical and Planetary Sciences, University of Colorado, 391 UCB, Boulder, CO 80309-0391, USA}
\email{taeho.ryu@colorado.edu}
\affiliation{Max-Planck-Institut für Astrophysik, Karl-Schwarzschild-Straße 1, 85748 Garching bei München, Germany}

\author[0000-0002-0716-3801]{Chen Wang}
\affiliation{School of Astronomy and Space Science, Nanjing University, Nanjing 210023, China}
\email{chen_wang@nju.edu.cn}
\affiliation{Max-Planck-Institut für Astrophysik, Karl-Schwarzschild-Straße 1, 85748 Garching bei München, Germany}

\author[0000-0003-3551-5090]{Alison Sills}
\affiliation{Department of Physics \& Astronomy, McMaster University, 1280 Main Street West, Hamilton, ON L8S 4M1, Canada}
\email{asills@mcmaster.ca}

\author[0000-0003-3308-2420]{Rüdiger Pakmor}
\affiliation{Max-Planck-Institut für Astrophysik, Karl-Schwarzschild-Straße 1, 85748 Garching bei München, Germany}
\email{rpakmor@mpa-garching.mpg.de}

\begin{abstract}
A significant fraction of stars experience close interactions, including collisions resulting from gravitational encounters and mergers within close binary systems. These processes can produce more massive stars that may give rise to relatively rare objects such as blue stragglers. Distinguishing the outcomes of collisions and mergers is challenging yet essential for interpreting observations. This study utilizes the magnetohydrodynamics code \texttt{AREPO} to simulate collisions and mergers of $5$ to $10 \Msun$ main-sequence stars, systematically comparing the properties of the resulting products. Both collisions and mergers yield more massive, strongly magnetized, rapidly and differentially rotating stars with cores enriched in hydrogen, but notable quantitative differences emerge. Merger products exhibit core hydrogen fractions up to $10\%$ higher than those of collision products. In both scenarios, turbulent mixing amplifies magnetic field energies by $9$ to $12$ orders of magnitude. However, magnetic fields in small-impact-parameter collision products display small-scale reversals that may dissipate over time, whereas merger products and large-impact-parameter collision products develop large-scale ordered, potentially long-lived magnetic fields. Additionally, only merger products display magnetically driven, bipolar outflows with radial velocities up to $300 \kmsinv$. These distinctions may result in different long-term evolutionary outcomes, which warrant further investigation in future studies.
\end{abstract}

\keywords{\uat{Stellar dynamics}{1596} --- \uat{Binary stars}{154} --- \uat{Stellar mergers}{2157} --- \uat{Stellar evolution}{1599} --- \uat{Hydrodynamics}{1963}}

\section{Introduction} \label{sec:intro}

Only some stars live out their lifetimes in isolation, while the rest interact with other stars \citep[e.g.,][]{2003ARA&A..41...57L,2013ARA&A..51..269D}. These interactions provide pathways for stars to gain mass via dynamically induced collisions \citep[e.g.,][]{1976ApL....17...87H,2025arXiv250918421S} and binary mergers triggered by unstable mass transfer \citep[e.g.,][]{1975ApJ...198..383L,2001ApJ...552..664N,2025arXiv250918421S}. Such interactions play a key role in shaping the stellar demographics in clusters and producing observable transients. Therefore, studying them is vital in understanding the evolution of star clusters and the formation of objects such as blue straggler stars \citep[BSSs; e.g.,][]{1953AJ.....58...61S,2025ARA&A..63..467M}, X-ray binaries \citep[e.g.,][]{1993ARA&A..31...93V,2006csxs.book..623T}, millisecond pulsars \citep[e.g.,][]{1982CSci...51.1096R,1991PhR...203....1B}, thermonuclear supernova progenitors \citep[e.g.,][]{1973ApJ...186.1007W,1984ApJ...277..355W,2025A&ARv..33....1R}, gravitational-wave progenitors \citep[e.g.,][]{2014LRR....17....3P,2017ApJ...846..170T,2022PhR...955....1M}, etc.

Stellar collisions are expected to occur frequently in the cores of dense star clusters where stellar densities are high enough to allow close encounters between single stars \citep[e.g.,][]{1976ApL....17...87H,1997ApJ...487..290S,2005MNRAS.358..716S}. In the densest clusters, such encounters result in runaway collisions \citep{1999A&A...348..117P,2004Natur.428..724P,2006MNRAS.368..141F}, possibly giving rise to very massive stars. Collisions also arise from binary-single and binary-binary interactions \citep[e.g.,][]{1993ApJ...415..631S,2004MNRAS.352....1F,2013ApJ...777..106C}, potentially dominating collisions in star clusters on account of their larger cross sections for interactions. As a result, the rates of stellar collisions are strongly correlated with binary fractions \citep{2013ApJ...777..106C}. Hydrodynamics studies of collisions in single-single \citep[e.g.,][]{1995ApJ...445L.117L,1996ApJ...468..797L,1997ApJ...487..290S,2001ApJ...548..323S,2005MNRAS.358.1133F,2013MNRAS.434.3497G,2024MNRAS.528.6193R,2025ApJ...980L..38R} and binary-single interactions \citep{2003MNRAS.345..762L,2010MNRAS.402..105G,2022MNRAS.516.2204R,2023MNRAS.519.5787R,2023MNRAS.525.5752R,2024MNRAS.527.2734R} have shown that the resulting products are rejuvenated owing to chemical mixing, puffed up, and fast rotators. Further evolution using stellar evolution codes \citep[e.g.,][]{1997ApJ...487..290S,2001ApJ...548..323S,2013MNRAS.434.3497G} reveals that their long-term evolution can be significantly influenced by their properties right after collision. Notably, while \citet{2025ApJ...980L..38R} conducted a magnetohydrodynamics (MHD) study of collisions of low-mass main-sequence (MS) stars, MHD simulations of massive star collisions remain largely unexplored. 

Close binary stars can undergo mass transfer interactions when one of them overfills their Roche lobe (e.g., \citealp{1975ApJ...198..383L,2022A&A...659A..98S}; see \citealp{2024ARA&A..62...21M} for a recent review) and can eventually merge \cite[e.g.,][]{2001ApJ...552..664N,2008MNRAS.384.1263C} if mass transfer becomes unstable from L$_2$-overflow \citep[e.g.,][]{1979ApJ...229..223S,1998CoSka..28..101P,2021A&A...650A.107M} or as a result of tidal instability \citep[e.g.,][]{1879RSPS...29..168D,1973ApJ...180..307C}. Mergers are preceded by a contact binary phase \citep[e.g.,][]{2024A&A...682A.169H} and may be observed as luminous red novae \citep[e.g.,][]{2003ApJ...582L.105S,2011A&A...528A.114T}. Massive contact binaries \citep[e.g.,][]{2008ApJ...681..554P,2010ApJ...709..632K,2013A&A...559A..22M,2014A&A...572A.110L,2014NewA...30..100O,2014NewA...31...32Y} are, hence, promising candidates for the mergers of massive stars. Additionally, binary hardening in clusters due to binary-single interactions \citep[e.g.,][]{1975MNRAS.173..729H,2006MNRAS.373.1188B} contributes to bringing stars closer, facilitating mergers. It is challenging to model stellar mergers using hydrodynamics because the thermal (or nuclear) timescales of mass transfer prior to merger are much longer than the dynamical timescales for which such studies are suited; nevertheless, hydrodynamics simulations of mergers of stars have been performed \citep[e.g.,][]{1995ApJ...438..887R,2014ApJ...786...39N,2017ApJS..229...27M,2024ApJ...962..168S}. \citet{2019Natur.574..211S} conducted the only MHD study of a stellar merger of two massive MS stars by employing a scheme to artificially accelerate the mass transfer phase. Analogous to collision products, merger products are extended, longer-lived, and fast-rotating.

Stellar collisions and mergers are among the proposed formation channels for BSSs (see \citealp{2024arXiv241010314W,2025ARA&A..63..467M} for recent reviews), those located above the MS turn-off points in the color-magnitude diagrams of open and globular star clusters (e.g., \citealp{1953AJ.....58...61S,1955ApJ...121..616J,1958ApJ...128..174B}). More recently, stars above the MS turnoff have also been observed in young stellar clusters \citep[e.g.,][]{2017ApJ...844..119L,2018ApJ...856...25L,2021A&A...652A..70B,2023A&A...672A.161M}, hinting at the presence of massive BSSs. The collision channel of BSS formation is believed to contribute more in very dense globular cluster cores \citep[e.g.,][]{2004MNRAS.349..129D}, while the binary merger channel tends to dominate in less dense open clusters \citep[e.g.,][]{2009Natur.462.1032M,2011Natur.478..356G}. Furthermore, detailed binary evolution models suggest that mergers can play an important role in the formation of (massive) BSSs in young open clusters \citep[e.g.,][]{2020ApJ...888L..12W,2022NatAs...6..480W}. 

In the field, stellar collisions and mergers are expected to occur frequently in triple- and quadruple-star systems \citep[e.g.,][]{2022ApJ...925..178H,2024arXiv241214022P,2025ApJ...978...47S,2025A&A...703A.123K}, where more than $50 \%$ of all stars, especially massive ones, reside \citep[e.g.,][]{2013ARA&A..51..269D,2017ApJS..230...15M,2023ASPC..534..275O}. For example, \citet{2024arXiv241214022P} found that $30 \%$ -- $40 \%$ of massive multiple-star systems undergo these interactions, primarily facilitated by significant orbital perturbations arising from (von Zeipel)-Lidov-Kozai oscillations (\citealp{1910AN....183..345V,1962P&SS....9..719L,1962AJ.....67..591K}; see \citealp{2016ARA&A..54..441N} for review). However, these field stars are challenging to identify observationally, since, unlike in star clusters, there are no other stars around to compare relative ages.

Stellar mergers and collisions may also pave the way to forming stars with strong magnetic fields \citep[e.g.,][]{2009MNRAS.400L..71F,2014MNRAS.437..675W,2016MNRAS.457.2355S,2025arXiv250918421S}, as suggested by previous MHD simulations \citep{2019Natur.574..211S,2025ApJ...980L..38R}. Such interactions in massive stars can explain the observations of $\sim 10 \%$ of O, B, and A stars having strong magnetic fields $\sim 10 \kG$ \citep[e.g.,][]{2009ARA&A..47..333D,2015A&A...582A..45F}, many of them being chemically peculiar Ap/Bp stars \citep[e.g.,][]{1947ApJ...105..105B,1992A&ARv...4...35L,2007A&A...475.1053A}. These massive magnetic stars can end their lives as magnetars, which have surface magnetic fields of $10^{13}$ -- $10^{15} \G$ \citep{1992ApJ...392L...9D,2014A&A...565A..90C} and are predominantly located in regions with massive OB stars such as star-forming regions and young open clusters \citep[e.g.,][]{2014ApJS..212....6O}.

Given the existence of astrophysical objects expected to form through both collisions and mergers and their anticipated similarities in mass growth and potential stellar magnetism, a critical question arises: can we clearly distinguish the resulting products of collisions and mergers? This question is particularly relevant in interpreting the origin of observed BSSs. Many observational studies on BSSs have primarily relied on host cluster properties, such as stellar density and dynamical evolution, to infer their formation mechanisms \citep[e.g.,][]{2012Natur.492..393F,2016ApJ...833L..29L} owing to the limited theoretical understanding of the similarities and differences between the outcomes of collisions and mergers. This motivates our study, which employs moving-mesh MHD to systematically compare the mixing, rotation, structures, and magnetic fields of the resulting products of collisions and mergers of massive MS stars. 

This paper is organized as follows. Section \ref{sec:method} outlines the methods we use to generate the initial conditions and perform our MHD simulations. Section \ref{sec:result} presents the comparative results of collisions and mergers. Section \ref{sec:conclude} discusses the implications and summarizes the conclusions.

\section{Methods} \label{sec:method}

\subsection{Magnetohydrodynamics}
We simulate collisions and mergers of stars using the 3D MHD code \arepo{} \citep{2010MNRAS.401..791S,2016MNRAS.455.1134P,2020ApJS..248...32W}. \arepo{} is an arbitrary Lagrangian-Eulerian, moving-mesh code that combines elements of the Eulerian finite-volume method and the Lagrangian smooth particle hydrodynamics method. Several studies have employed the \arepo{} code to simulate phenomena involving stars, such as tidal disruption events \citep[e.g.,][]{2023MNRAS.519.5787R,2023MNRAS.525.5752R,2024MNRAS.527.2734R,2024A&A...685A..45V}, collisions of stars \citep[e.g.,][]{2024MNRAS.528.6193R,2025ApJ...980L..38R}, mergers of stars \citep[e.g.,][]{2019Natur.574..211S} and compact objects \citep[e.g.,][]{2013ApJ...770L...8P,2021MNRAS.503.4734P,2021A&A...649A.155G,2023MNRAS.523..527B,2024MNRAS.528.1906L,2024A&A...681A..41M,2025ApJ...988..184G}, and common envelope evolution \citep[e.g.,][]{2016ApJ...816L...9O,2016MNRAS.462L.121O,2020A&A...644A..60S,2020A&A...642A..97K,2022A&A...660L...8O,2025A&A...698A.133V}.

\arepo{} solves the Euler MHD equations on an unstructured Voronoi mesh that moves with (roughly) the local velocity of the fluid and computes the fluxes between cell interfaces using the HLLD Riemann solver (\citealp{2005JCoPh.208..315M}; see \citealp{2011MNRAS.418.1392P} for implementation). Self-gravity is included, and the Powell cleaning scheme (\citealp{1999JCoPh.154..284P}; see \citealp{2013MNRAS.432..176P} for implementation) is employed to control the magnetic field divergence. Refinement is triggered when adjacent cells have volume differences of more than a factor of ten; mass refinement is not enabled. Finally, we set the minimum gravity softening factor to be one-tenth the size of the smallest cell.

\subsection{Stellar models}

\begin{figure*}[ht!]
\centering
\includegraphics[width=0.48\textwidth]{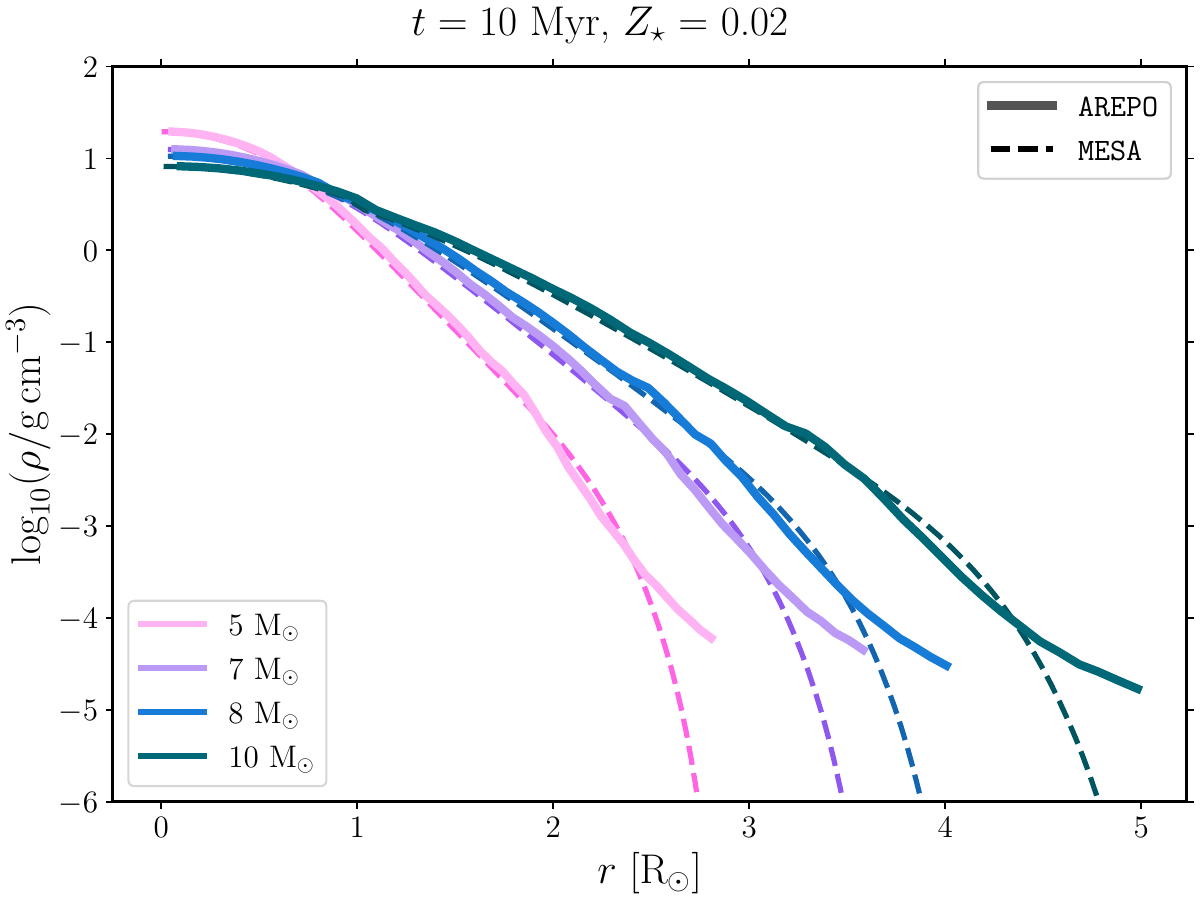}
\includegraphics[width=0.48\textwidth]{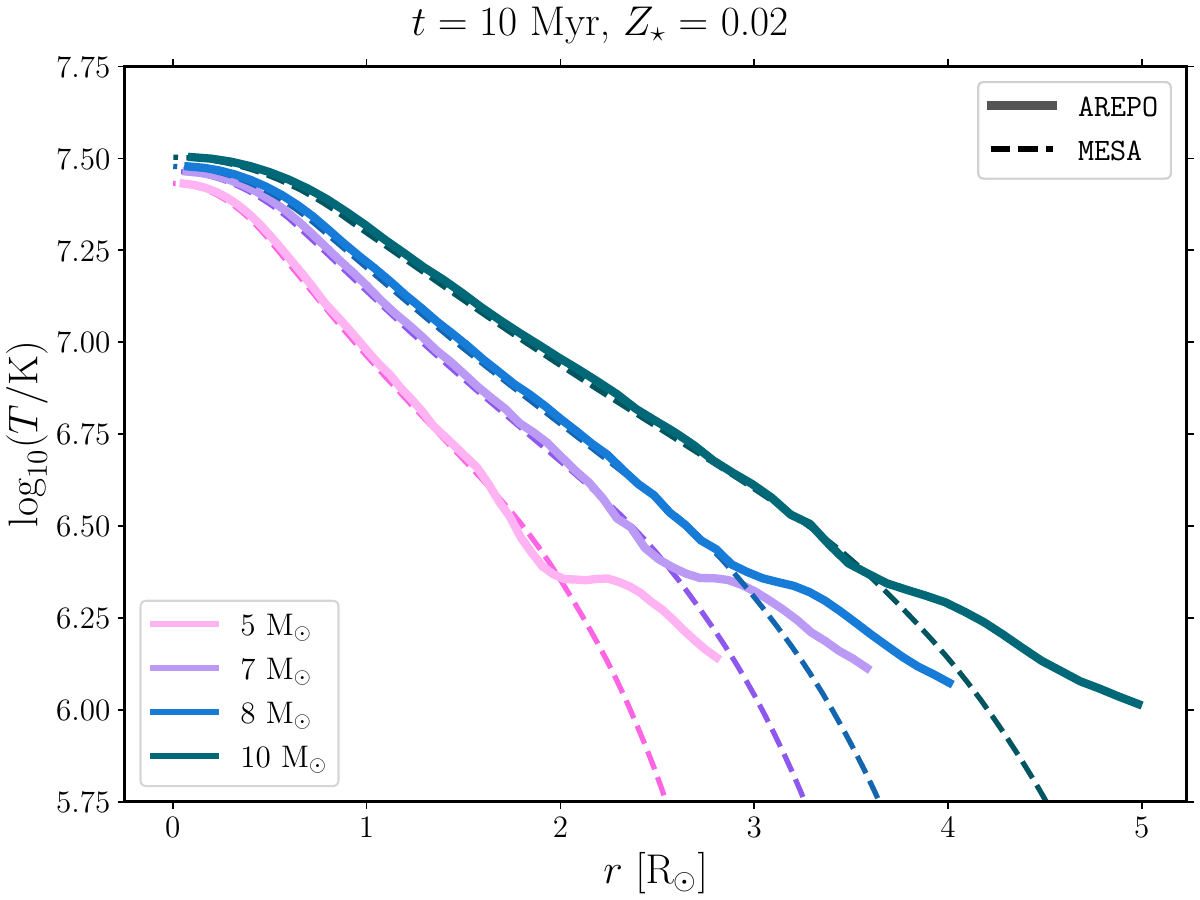}
\caption{Initial density and temperature profiles of $10 \Myr$-old MS stars of different masses, corresponding to Table \ref{tab:star_models}. The thinner dashed and thicker solid lines represent the 1D profiles from \mesa{} and 3D shell-averaged and relaxed profiles from \arepo{}, respectively. The \arepo{} profiles diverge from their \mesa{} counterparts at densities of $\lesssim 10^{-4} \gcmcbinv$  and temperatures of $\lesssim 10^{6.25} \K$. \label{fig:den_temp_init}}
\end{figure*}

We employ the 1D stellar evolution code \mesa{} \citep{2011ApJS..192....3P} to generate nonrotating stellar models of $10 \Myr$-old MS stars of metallicities $\Zs = 0.02$ and masses ranging from $5 \Msun$ to $10 \Msun$\footnote{The \texttt{MESA} star inlists are available at \dataset[doi: 10.5281/zenodo.18249784]{10.5281/zenodo.18249784}.}. These parameter ranges probe the early evolution of stars in young stellar clusters. Following \citet{2024A&A...685A..45V}, we convert these 1D profiles of densities, pressures, internal energies, and chemical abundances to 3D \arepo{} cells using the HEALPix-shell-layering procedure \citep{2005ApJ...622..759G} from \citet{2017A&A...599A...5O} and calculate the thermodynamic variables of each cell from the \texttt{Helmholtz} equation of state \citep{2000ApJS..126..501T}. We then relax each of the \arepo{} stars in isolation for five stellar dynamical timescales $\tdyn = (\rs^3 / G \ms)^{0.5}$, where $\ms$ and $\rs$ are the star's mass and radius, respectively, to ensure that they reach a steady state.  Table \ref{tab:star_models} shows the \mesa{}-derived parameters of the stellar models we consider, with the stellar thermal timescale given by $\tth = G \ms^2/\rs \Ls$, where $\Ls$ is the star's luminosity provided by \mesa{}, and the MS lifetime $\tms$ representing the age when the core hydrogen fraction $\Xc < 10^{-3}$.

\begin{deluxetable*}{ccccccccc}[ht!]
\tablecaption{Parameters of $10 \Myr$-old MS stellar models with metallicities $\Zs = 0.02$. \label{tab:star_models}}
\tablehead{
\colhead{$\ms$ [$\Msun$]} & \colhead{$\Ncells$ [$\times 10^6$]} & \colhead{$\Xc$} & \colhead{$\rs$ [$\Rsun$]} & \colhead{$\Ls$ [$\Lsun$]} & \colhead{$\tdyn$ [$\hr$]} & \colhead{$\tth$ [$\kyr$]} & \colhead{$\tms$} [$\Myr$]
}
\startdata
$5.0$ & $1.0$ & $0.66$ & $2.81$ & $559$ & $0.93$ & $500$ & $99.1$ \\
$7.0$ & $1.4$ & $0.62$ & $3.58$ & $1945$ & $1.13$ & $221$ & $45.9$ \\
$8.0$ & $1.6$ & $0.59$ & $4.01$ & $3201$ & $1.26$ & $157$ & $34.7$ \\
$10.0$ & $2.0$ & $0.51$ & $4.98$ & $7465$ & $1.55$ & $85$ & $22.7$ \\
\enddata
\tablecomments{The columns represent the star's mass, approximate number of \arepo{} cells, core hydrogen fraction, radius, luminosity, dynamical timescale, thermal timescale, and MS lifetime.}
\end{deluxetable*}

The simulations are carried out using the ideal MHD equations that assume zero electrical resistivity $\eta$, and consequently zero magnetic diffusivity $\eta\,c^{2}/ 4 \pi$ (Gaussian units). This assumption of a large magnetic Reynolds number $\mathrm{Rm} \propto 1/\eta$ is largely reasonable for stellar plasmas. Therefore, the effects of resistivity and viscosity (also zero in ideal MHD) in our simulations are purely numerical.

Intermediate- and high-mass stars, which have convective cores and radiative envelopes, are observed to have a wide range of magnetic field strengths \cite[e.g.,][]{2009ARA&A..47..333D,2012SSRv..166..145W}, with only a small minority having strong magnetic fields. Given the uncertainties in field configurations, we initialize simple dipole magnetic fields $\bm{B}$ centered around each star:
\begin{equation}
    \bm{B}(\bm{r}) = \frac{\Bsurf}{2} \frac{3 \bm{\hat{r}} (\bm{\hat{r}} \cdot \bm{\hat{m}}) - \bm{\hat{m}}}{(|\bm{r}|/\rs)^3}
\end{equation}
Here, $\bm{r}$ is the radius vector with respect to the star's center of mass (CoM), $\Bsurf$ is the surface magnetic field at the poles,  and $\bm{\hat{m}}$ is the unit vector along the dipole axis. We choose $\Bsurf = 1\,\mathrm{G}$, and $\bm{\hat{m}}$ to be perpendicular to the orbital plane. Furthermore, we limit the initial magnetic field strength to $100\,\Bsurf$ to avoid very large field strengths at the core. The impact of the initial field configuration and orientation is expected to be small (see \citealp{2025ApJ...980L..38R}) because of the turbulent motion of the stellar material (see Section \ref{sec:result}) during coalescence (henceforth used to collectively refer to both collisions and mergers).

We perform resolution tests to ensure that our chosen cell resolutions $\Ncells$ produce convergent results of magnetic field amplification in collisions and mergers. Checking for convergence in the magnetic energy $\displaystyle \EB = \sum_{\mathrm{cells}} (|\bm{B}_{\mathrm{cell}}|^{2} V_{\mathrm{cell}}) / 8\pi$ (Gaussian units), where $V_{\mathrm{cell}}$ is the cell volume, we opt for a final resolution of $\sim 2 \times 10^5$ cells per $\Msun$\footnote{Approximate because the refinement of cells in \arepo{} results in a slight increase in $\Ncells$ every time step.} (see the second column of Table \ref{tab:star_models}). For this resolution, the numerical Reynolds number \citep[Equation 2 in][]{2024MNRAS.528.2308P} $\mathrm{Re}_{\mathrm{num}} \approx 3 \mathcal{L}/d_{\mathrm{cell}} \sim 700$ -- $1000$. Here, $\mathcal{L} \sim 0.5\, \rtot$, with $\rtot = \rsone + \rstwo$, is the turbulent injection scale and $d_{\mathrm{cell}}$ is the average diameter of the 100 smallest \arepo{} cells. $\mathrm{Re}_{\mathrm{num}} \gtrsim 100$ is necessary for the magnetic field amplification not to be suppressed by numerical dissipation. 

Figure \ref{fig:den_temp_init} shows the density and temperature profiles of the four stellar models indicated in Table \ref{tab:star_models}. The \mesa{} and \arepo{} profiles agree very well and diverge only at very low densities near the stellar surfaces because of unresolved gradients in \arepo{}. These regions that deviate from the \mesa{} models only represent fractional masses less than $10^{-3}$ and therefore should not appreciably affect the dynamics.

\begin{deluxetable*}{cc cc cccc cc ccc}[ht!]
\tablecaption{Initial parameters of the collision and merger simulations and the final properties of their resultant products $\sim 60\,\tdynrtot$ after coalescence. \label{tab:new_properties}}
\tablehead{
\multicolumn{2}{c}{} & \nocolhead{} & \nocolhead{} & \multicolumn{4}{c}{Collisions} & \nocolhead{} & \nocolhead{} & \multicolumn{3}{c}{Mergers} \\
\colhead{$\msone$} & \colhead{$\mstwo$} & \nocolhead{} & \nocolhead{} & \colhead{$b$} & \colhead{$\mbound$} & \colhead{$\Etot$} & \colhead{$\Lax$} & \nocolhead{} & \nocolhead{} & \colhead{$\mbound$} & \colhead{$\Etot$} & \colhead{$\Lax$} \\
\vspace{-0.6cm} \\
\colhead{[$\Msun$]} & \colhead{[$\Msun$]} & \nocolhead{} & \nocolhead{} & \colhead{} & \colhead{[$\Msun$]} & \colhead{[$\erg$]} & \colhead{[$\gcmcbinv$]} & \nocolhead{} & \nocolhead{} & \colhead{[$\Msun$]} & \colhead{[$\erg$]} & \colhead{[$\gcmcbinv$]}
}
\startdata
$10.0$ & $8.0$ & & & $0.25$ & $17.61$ & $-3.56 \times 10^{50}$ & $2.26 \times 10^{53}$ & & & $17.94$ & $-4.14 \times 10^{50}$ & $3.05 \times 10^{53}$ \\
$10.0$ & $7.0$ & & & $0.25$ & $16.55$ & $-3.34 \times 10^{50}$ & $1.97 \times 10^{53}$ & & & $16.91$ & $-3.64 \times 10^{50}$ & $2.86 \times 10^{53}$ \\
$10.0$ & $5.0$ & & & $0.25$ & $14.55$ & $-2.83 \times 10^{50}$ & $1.46 \times 10^{53}$ & & & $14.68$ & $-3.09 \times 10^{50}$ & $1.93 \times 10^{53}$ \\
$8.0$ & $7.0$ & & & $0.25$ & $14.69$ & $-2.84 \times 10^{50}$ & $1.63 \times 10^{53}$ & & & $14.96$ & $-3.22 \times 10^{50}$ & $2.30 \times 10^{53}$ \\
$8.0$ & $5.0$ & & & $0.25$ & $12.68$ & $-2.44 \times 10^{50}$ & $1.17 \times 10^{53}$ & & & $12.88$ & $-2.66 \times 10^{50}$ & $1.65 \times 10^{53}$ \\
$7.0$ & $5.0$ & & & $0.25$ & $11.77$ & $-2.19 \times 10^{50}$ & $1.05 \times 10^{53}$ & & & $11.93$ & $-2.45 \times 10^{50}$ & $1.46 \times 10^{53}$ \\
$10.0$ & $5.0$ & & & $0.50$ & $14.58$ & $-2.84 \times 10^{50}$ & $1.93 \times 10^{53}$ & & & - & - & - \\
$8.0$ & $7.0$ & & & $0.50$ & $14.69$ & $-2.85 \times 10^{50}$ & $2.21 \times 10^{53}$ & & & - & - & - \\
\enddata
\tablecomments{The columns represent the two stars' initial masses, the collision impact parameters, and the bound masses, total energies, and spin angular momenta about the rotation axes for all the collision and merger products.}
\end{deluxetable*}

\subsection{Initial conditions}
With relaxed stellar models in hand, we run a suite of simulations of collisions and mergers. In addition to the MHD quantities, we follow the advection of two passive scalars $K_1$ and $K_2$ corresponding to the fluid cells of the two stars. $K_1$ and $K_2$ are scalar quantities initialized as `1' for the cells that make up the heavier and lighter stars, respectively, and `0' for all other cells. We define an additional passive scalar $K$ as a combination of $K_1$ and $K_2$:
\begin{equation}    
    \label{eq:passScal}
    K = \begin{cases}
        K_1 - K_2 &, \text{if } \rho > 10^{-5} \gcmcbinv \\
        0 &, \text{if } \rho \le 10^{-5} \gcmcbinv
        \end{cases}
\end{equation}
For cells with densities greater than the threshold given above, $K_1 + K_2 \approx 1$. This is not true for very low density cells, since the `vacuum' cells initially have $K_1 = K_2 = 0$. Thus, $K$ quantifies the mixing through the course of each interaction, with $K > 0$ (or $K_1 > K_2$) indicating a higher fraction of the heavier star and $K < 0$ (or $K_1 < K_2$) indicating a higher fraction of the lighter star. 

The masses and, consequently, the mass ratios are varied while keeping the magnetic field strengths and directions constant. Table \ref{tab:new_properties} lists the initial masses of the 14 simulations: 6 collisions with a lower impact parameter, 2 collisions with a higher impact parameter, and 6 mergers. Defining the dynamical timescales at $\rtot$ as $\tdynrtot = (\rtot^3 / G \ms)^{0.5}$, all simulations are run up to $\sim 60\,\tdynrtot$ post-coalescence (5 to 8 days). The specific details of the two interaction channels are described below.

\begin{figure*}[ht!]
\centering
\includegraphics[width=\textwidth]{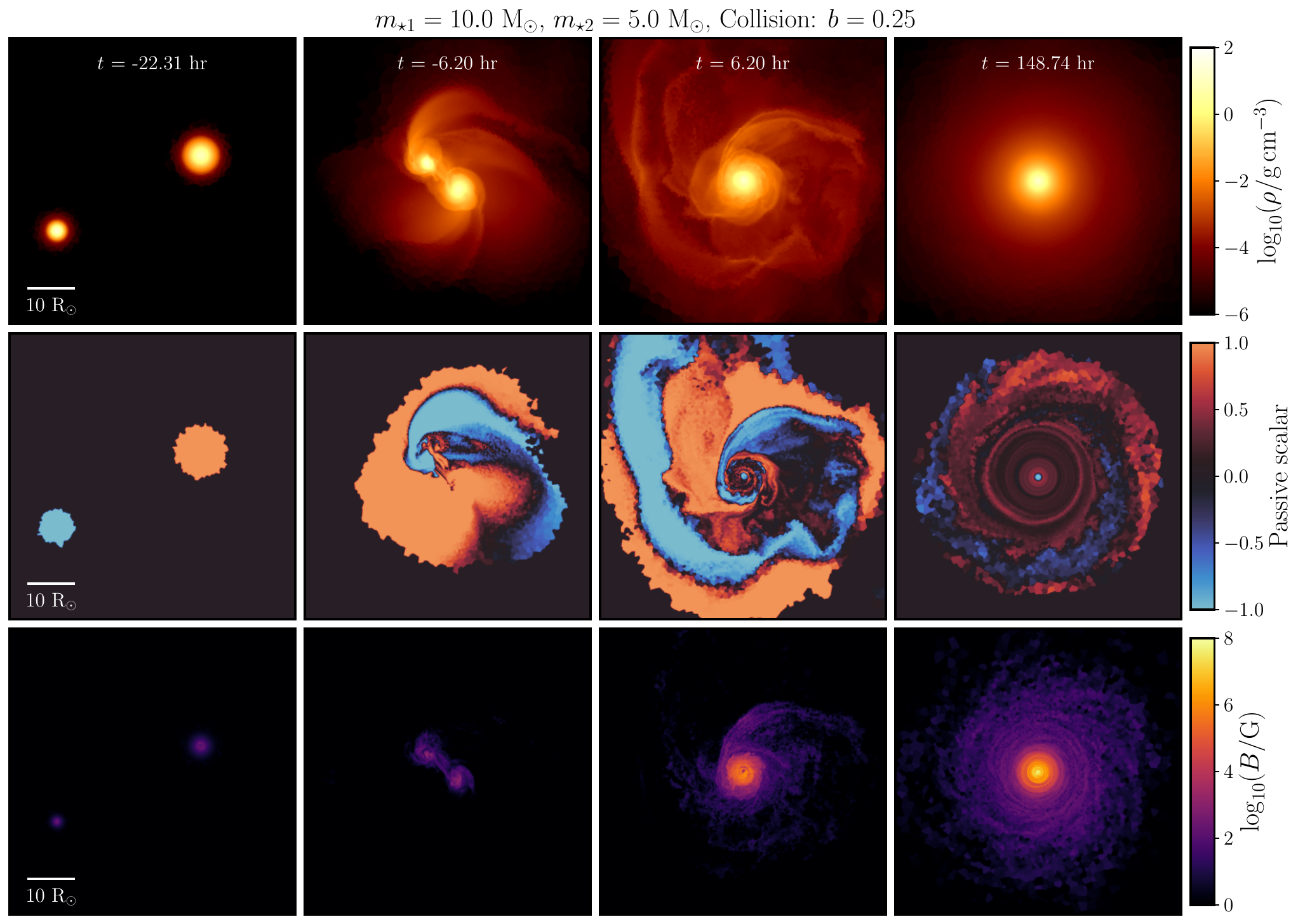}
\caption{Mid-plane snapshots of a stellar collision between a $10 \Msun$ star and a $5 \Msun$ star when $b = 0.25$, $10 \Myr$ into their MS lifetimes. The three rows illustrate slices, in the collision plane, of densities $\rho$ (top), passive scalars $K$ (middle), and magnetic field magnitudes $B$ (bottom). The four columns illustrate these quantities at the times $t$ before collision (first column), just before collision (second column), just after collision (third column), and after $\sim 60\,\tdynrtot$ following collision (fourth column). The core of the collision product is dominated by the initially less massive star, while the outer layers are composed of material from both stars in varying fractions. Furthermore, the amplification of the magnetic field is clearly visible across the panels. \label{fig:coll_collage_evo}}
\end{figure*}

\begin{figure*}[ht!]
\centering
\includegraphics[width=\textwidth]{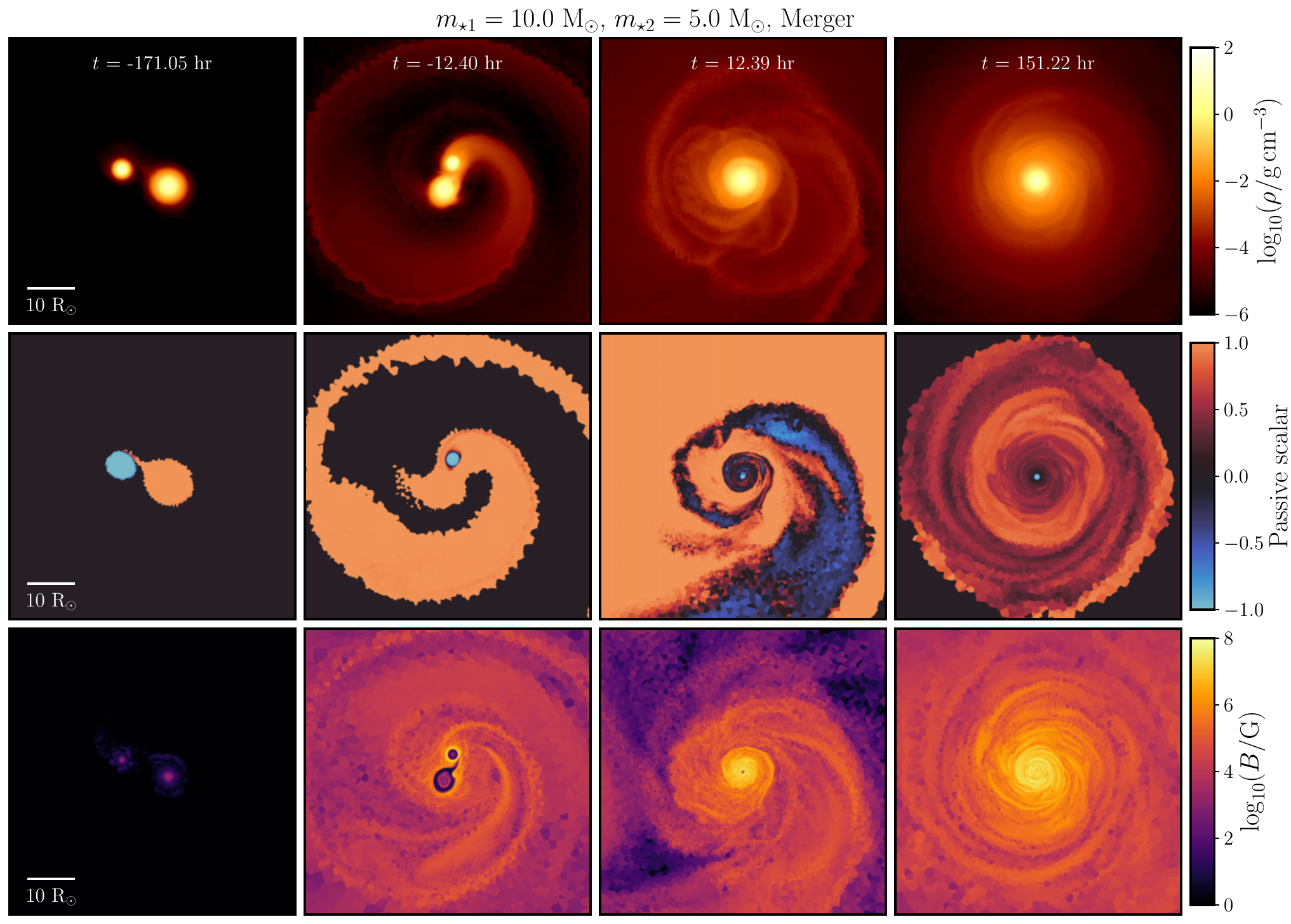}
\caption{Mid-plane snapshots, similar to Figure \ref{fig:coll_collage_evo}, of a stellar merger between a $10 \Msun$ star and a $5 \Msun$ star, $10 \Myr$ into their MS lifetimes. The merger product has a central region dominated by the initially less massive star, an outer region dominated by the initially more massive star, and a transition region composed of material from both stars (see Figure \ref{fig:coll_collage_evo} for comparison with the collision product). Moreover, the magnetic field within the merger product is notably higher. \label{fig:mer_collage_evo}}
\end{figure*}

\subsubsection{Collisions}
We assume the orbits in the collision scenario to be parabolic. This is a reasonable assumption since young stellar clusters are observed to have velocity dispersions $\sigma \sim 10$ -- $20 \kmsinv$ \citep[e.g.,][]{2010MNRAS.402.1750G}, yielding near-parabolic eccentricities\footnote{For a hyperbolic orbit, $e = 1 + r_\mathrm{p} v_{\infty}^2 / G M$, where  $r_\mathrm{p}$ is the periapsis distance, $v_{\infty}^2 \sim \sigma$ is the velocity at infinity, and $M$ is the total mass.} of $|e-1| \sim 10^{-4}$ for our typical stellar masses and radii. It should be mentioned that collisions are not very frequent in young stellar clusters; nonetheless, we consider these parameters for comparative purposes. We define the collision impact parameter as $b = \rp/\rtot$, where $\rp$ is the closest approach distance between the CoMs of the two stars. In 6 of the 8 collision simulations, we set $b = 0.25$. In the other 2 simulations, we set $b = 0.50$ to compare the effect of the impact parameter on the collision product. We initially position the relaxed stars $5\,\rtot$ apart prior to the pericenter passage.

\subsubsection{Mergers}
Unlike collisions, in reality, mergers occur on much longer timescales. For computational feasibility, we speed up the orbit shrinkage phase by adding artificial deceleration terms (similar to \citealp{2013ApJ...770L...8P,2021MNRAS.503.4734P}). The equations below, which are the same as Equation 2 in \citet{2021MNRAS.503.4734P}\footnote{Note that there is an error in that paper, which is corrected in our equation.}, quantify the antiparallel, azimuthal acceleration terms that result in an inspiral at a constant rate $v_\mathrm{insp}$:
\begin{equation}
    \label{eq:inspiral}
    f(r) = \begin{cases}
        \dfrac{Gv_{\mathrm{insp}}}{2 r^2\,(\msone + \mstwo)} &, \text{if } r \ge \rtot \\
        \quad 0 &, \text{if } r < \rtot
        \end{cases}
\end{equation}
\begin{subequations}
\begin{eqnarray}
    \bm{a}_{\star 1} = - \dfrac{\mstwo^2 \bm{v}_{\star 1}}{v_{\star 1}^2} f(r) \\
    \bm{a}_{\star 2} = - \dfrac{\msone^2 \bm{v}_{\star 2}}{v_{\star 2}^2} f(r)
\end{eqnarray}
\end{subequations}
Here, $r$ is the instantaneous distance between the two stars, $\bm{v}_{\star}$ and $\bm{a}_{\star}$ represent the star's velocity and acceleration vectors, respectively, and $v_\mathrm{insp} = 10 \kmsinv$ is the inspiral velocity we choose. In comparison, typical Keplerian circular velocities for the masses and distances considered are $\sim 500$ -- $600 \kmsinv$. This deviation from the circular velocity results in a slightly eccentric inspiral with $e \approx 0.1$. For a constant value of $v_\mathrm{insp}$, stars with a more skewed mass ratio merge more quickly (see Section \ref{sec:result} and Figure \ref{fig:mag_energy_tot}).

We position the relaxed (nonrotating) stars in a circular binary orbit with an initial semimajor axis of $1.5\,\rtot$ and run the simulations with the sped-up inspiral. As a comparison, using the Roche lobe radius approximation formula from \citet{1983ApJ...268..368E}, Roche lobe filling ensues at $\lesssim 1.31\,\rtot$ for our range of mass ratios (see Table \ref{tab:star_models}).

\section{Results} \label{sec:result}

In this section, we use the shorthand notation `$\msone$ + $\mstwo$' to denote an interaction between stars of masses $\msone$ and $\mstwo$ (in units of $\Msun$). We focus on comparing key properties of the collision and merger products, including chemical mixing, rotation, magnetic fields, and outflows. While these quantities are intrinsically at least 2D, we discuss them using their mass-weighted averages over equipressure surfaces as representative values\footnote{\mesa{} handles stellar rotation in 1D in a similar manner, with pressure being the independent coordinate.}. In addition to the averages, we indicate the spread in each calculated quantity within the equipressure surfaces by reporting the 15$^{\mathrm{th}}$ and 85$^{\mathrm{th}}$ percentiles. In the calculation of the averages and spreads, we only consider cells that are bound to the CoM of the new star, which we calculate iteratively as follows. We first find the initial values of the CoM position $\bm{r}_{\mathrm{CoM}}$ and velocity $\bm{v}_{\mathrm{CoM}}$ of all cells. We then calculate the total energies of all cells with respect to the CoM by summing up their kinetic, potential, and internal energies. Cells with negative total energies are bound to the initial CoM. In the next iteration, we calculate the new values of $\bm{r}_{\mathrm{CoM}}$ and $\bm{v}_{\mathrm{CoM}}$ of only the bound cells. This process is repeated until the CoM values do not vary.

\begin{figure*}[ht!]
\centering
\includegraphics[width=0.48\textwidth]{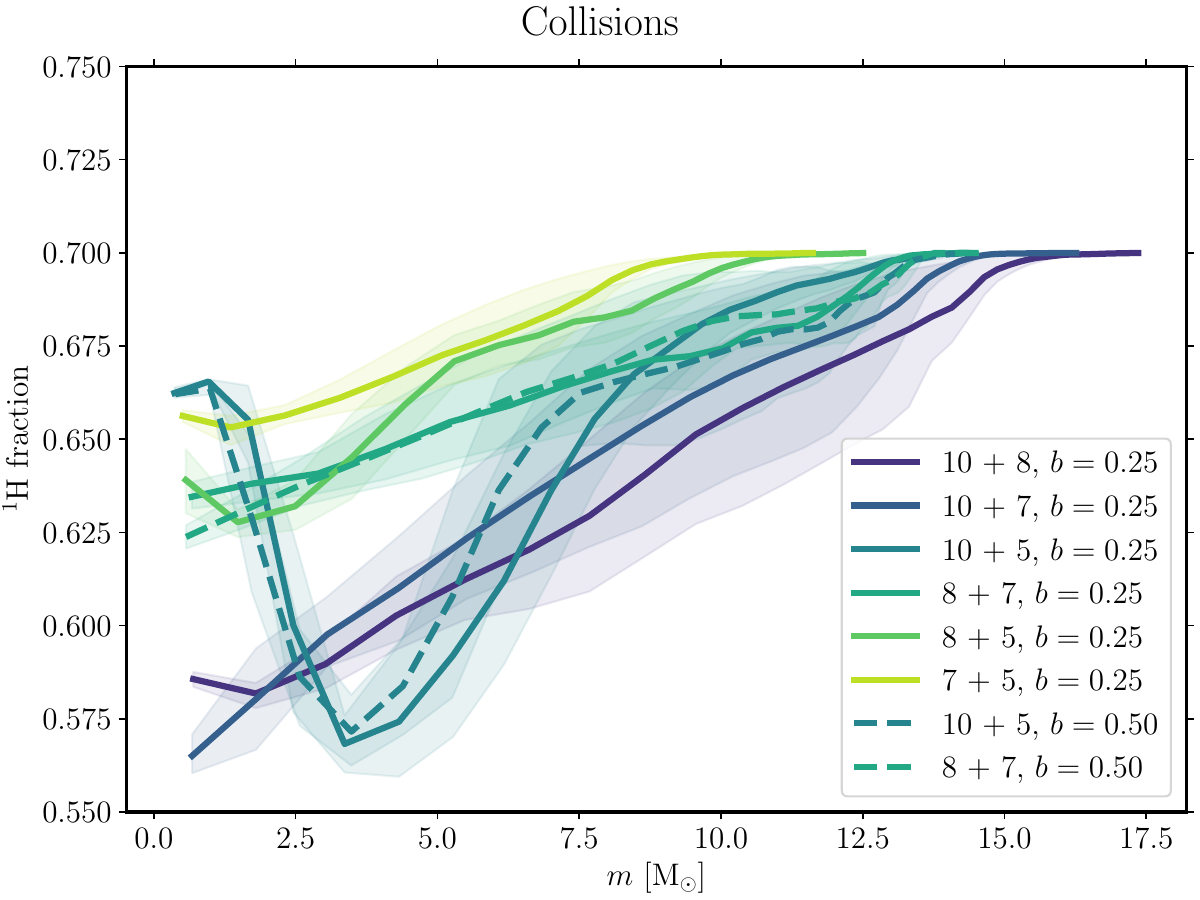}
\includegraphics[width=0.48\textwidth]{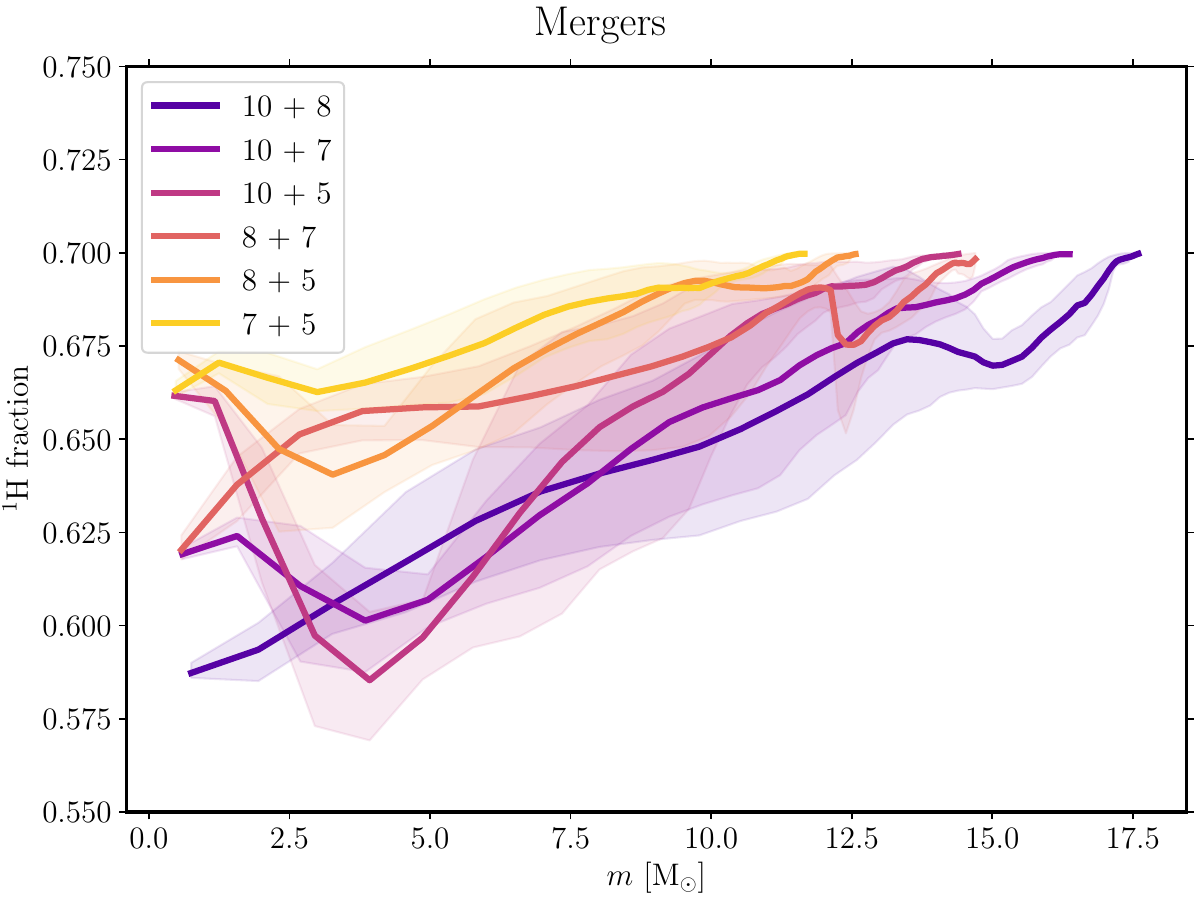}
\caption{Mass profiles of $^{1}\mathrm{H}$ fractions $X$ of collision (left) and merger (right) products for different masses. The dashed and solid lines in the collision plot represent $b = 0.50$ and $b = 0.25$, respectively. The shaded regions' boundaries represent the 15th and 85th percentiles. The resulting stars follow similar chemical profiles of $X$. As a comparison, Table \ref{tab:star_models} shows the core hydrogen fractions of the initial stars. \label{fig:profile_hydrogen}}
\end{figure*}

\subsection{Overview} \label{sec:overview}
During a collision event, the two stars go through multiple close passages and ultimately coalesce. During a merger event, two stars in a binary advance closer to each other owing to orbit shrinkage, undergo mass transfer and contact, and eventually coalesce into a single object. 

As an overview, we show successive moments of the 10 + 5, $b = 0.25$ collision and the 10 + 5 merger in Figures \ref{fig:coll_collage_evo} and \ref{fig:mer_collage_evo}, respectively. All figures display midplane slices of three quantities: density $\rho$, passive scalar $K$, and magnetic field $B$, at four different times: well before, just before, just after, and well after coalescence. The middle panels indicate that the cores of both the collision and merger products are predominantly made up of material from the less massive stars ($5 \Msun$). In other collision cases with different masses, however, collision products are not completely dominated by the initially less massive star. The mixing in the outer layers of the collision and merger products differs, but since the outer layers of both stars are of similar chemical compositions (see Section \ref{sec:mixing}) and contribute less mass compared to the cores, this is not expected to affect the stars' long-term evolution. Finally, both collision and merger products have significantly higher magnetic field strengths than their initial stars, as seen in the bottom panels. However, it is also clear that the magnetic field amplification in the merger product is considerably larger than in the $b = 0.25$ collision product (the color scales are the same). This is also true for other $b = 0.25$ collision and merger products. However, the collision simulations with $b = 0.50$ exhibit magnetic field amplification akin to mergers (see Section \ref{sec:magfield}).

Table \ref{tab:new_properties} shows the bound masses $\mbound$, total energies $\Etot$, and spin angular momenta $\Lax$ about the rotation axes of the collision and merger products after $\sim 60\,\tdynrtot$. We estimate the initial orbital energies and angular momenta of the mergers by assuming circular orbits with semimajor axes $\rtot$, where we stop the artificial inspiral (see Equation \ref{eq:inspiral}). More than $97 \%$ of the total masses remain bound to the resulting products. The total energies of collision products and merger products are within $1 \%$ -- $3 \%$ and $6 \%$ -- $13 \%$, respectively, of the summations of the initial orbital energies (relatively negligible for collisions) and the total energies of the initial stars. The spin angular momenta of all resulting products are significant fractions ($62 \%$ -- $68 \%$) of the orbital angular momenta of the initial systems (see also Section \ref{sec:rotation}).

\begin{figure*}[ht!]
\centering
\includegraphics[width=0.48\textwidth]{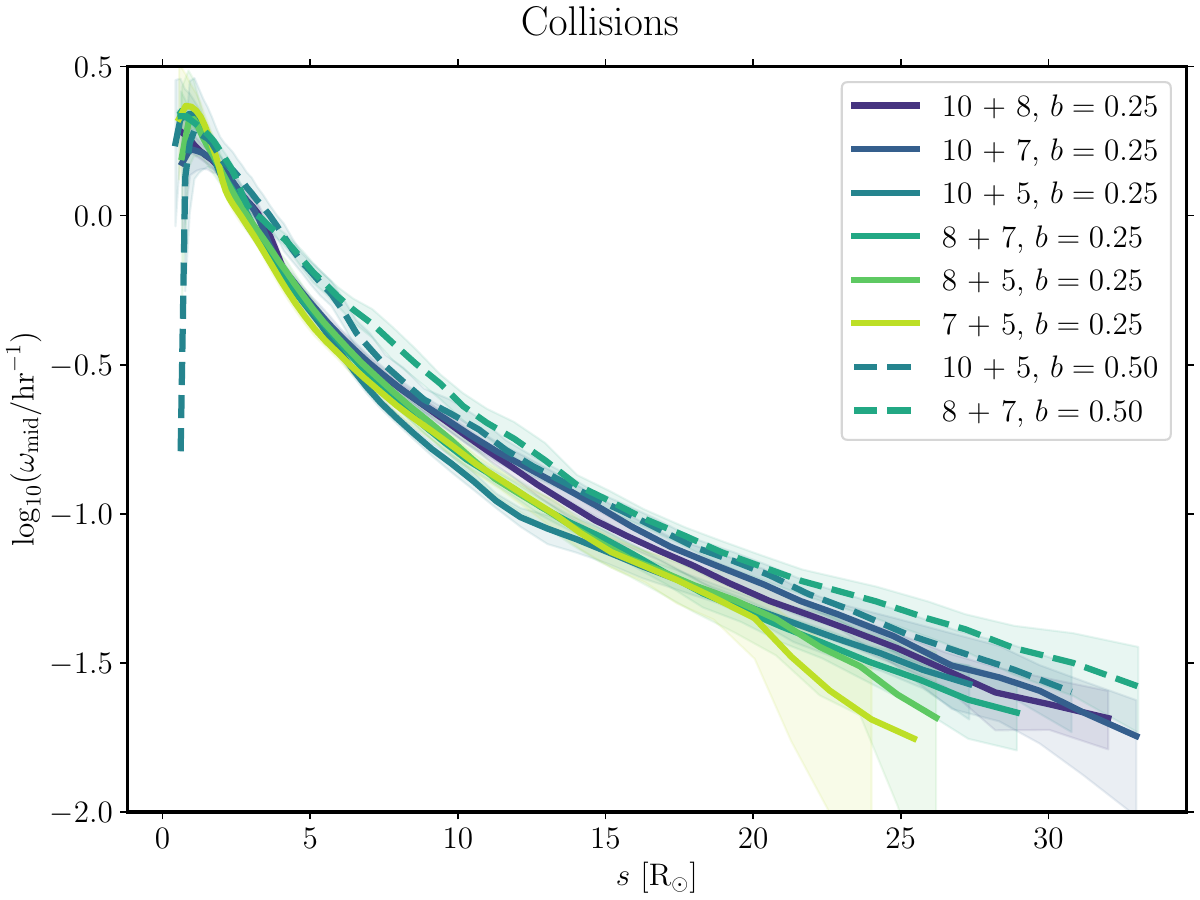}
\includegraphics[width=0.48\textwidth]{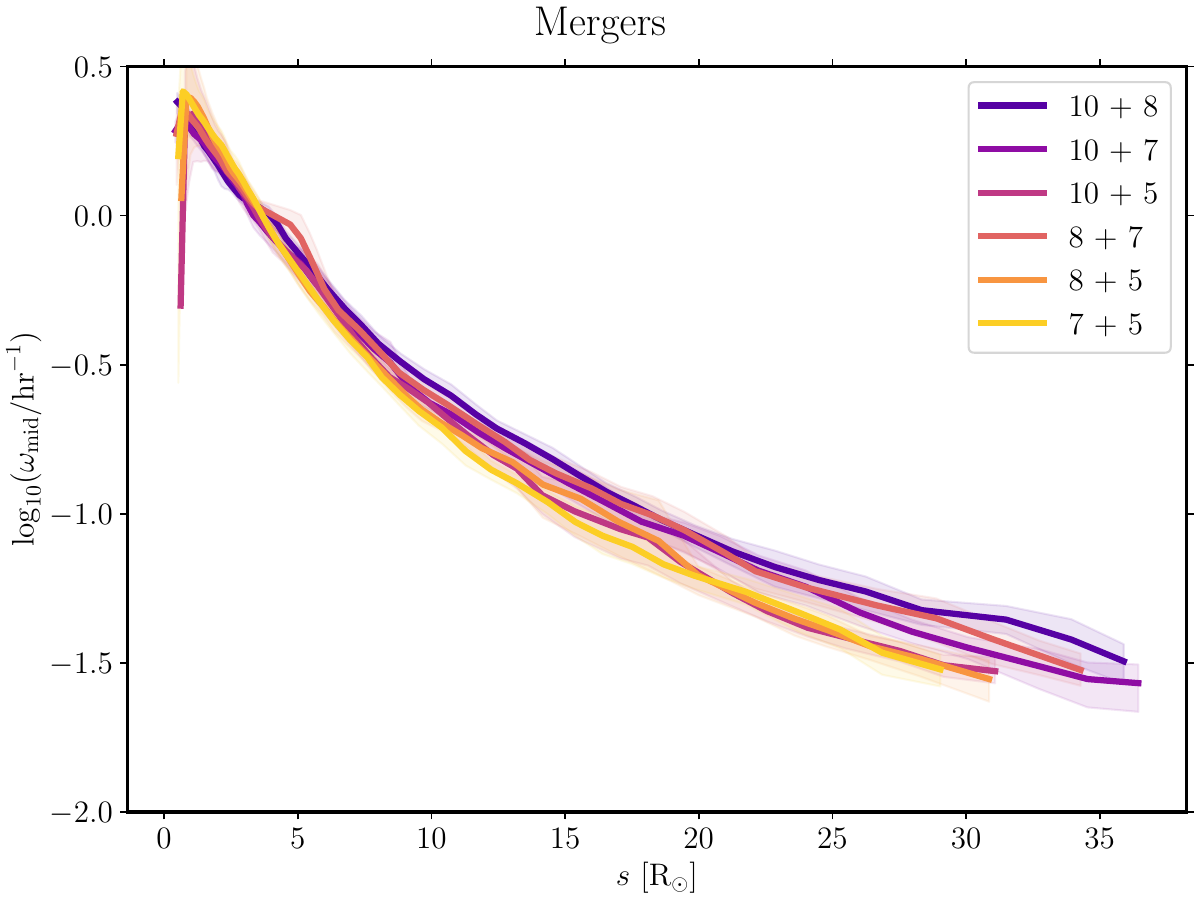}
\caption{Profiles of average angular frequencies in the midplanes (see text for definition) $\omegamid$ as functions of cylindrical radial distances $s$. The left and right panels represent collision and merger products, respectively, of different masses. Both collision and merger products have sharply declining angular velocity profiles. \label{fig:profile_angVels}}
\end{figure*}

\begin{figure*}[ht!]
\centering
\includegraphics[width=0.48\textwidth]{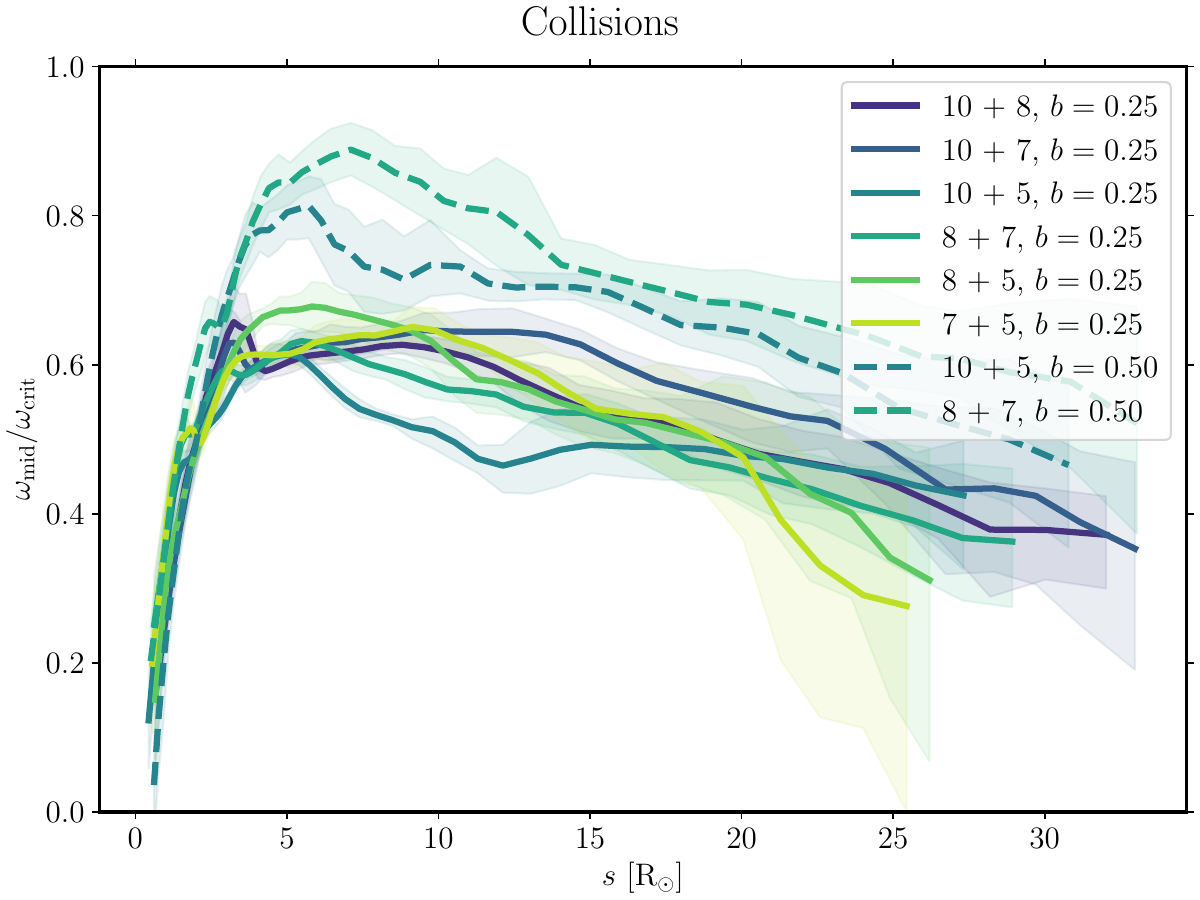}
\includegraphics[width=0.48\textwidth]{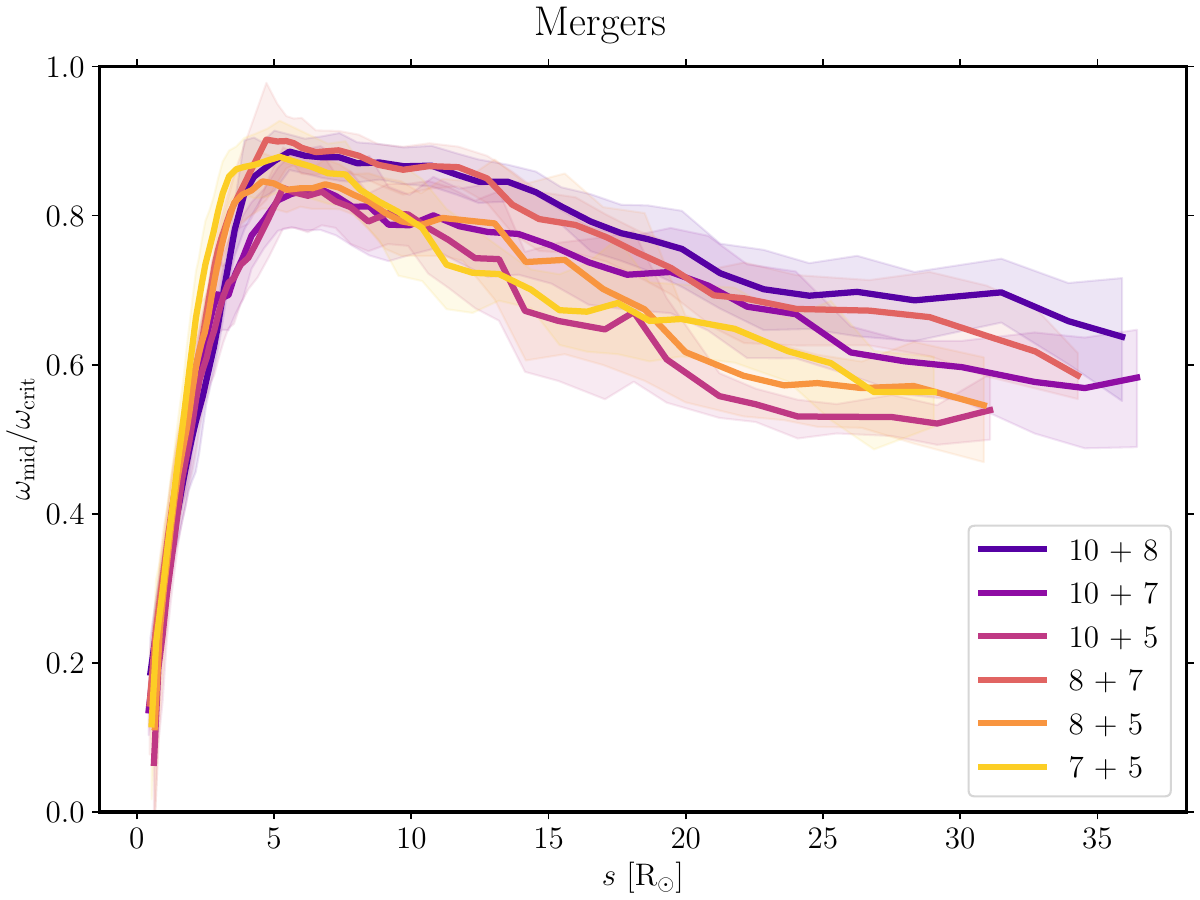}
\caption{Profiles of average angular frequency ratios in the midplanes $\omegamid/\omegacrit$ as functions of cylindrical radial distances $s$. The left and right panels represent collision and merger products, respectively, of different masses. Merger products rotate faster than collision products and are closer to breakup velocity. \label{fig:profile_angVel_ratios}}
\end{figure*}

\subsection{Chemical mixing} \label{sec:mixing}

Collision and merger interactions result in mixing in the stars, transporting helium from the cores to the new envelope and hydrogen from the envelopes to the new core, thus rejuvenating the star \citep[see also][]{1996ApJ...468..797L,2016MNRAS.457.2355S}. Figure \ref{fig:profile_hydrogen} shows the profiles of $^{1}\mathrm{H}$ (hydrogen) fractions $X$ for all collision and merger products. Since all stars are in the MS and have $\Zs = 0.02$ (see Section \ref{sec:method}), the $^{4}\mathrm{He}$ (helium) fractions $Y$ are inverses of $X$ and follow similar trends. The sorting of stellar layers in the resultant products depends on their relative entropies (or 'buoyancy' parameters; e.g., \citealp{1996ApJ...468..797L,2008MNRAS.383L...5G}, i.e., less buoyant material sinks to the new stellar core. For massive stars, the buoyancy parameters depend on densities, gas pressures, and radiation pressures \citep{2008MNRAS.383L...5G}. For an updated approach to entropy sorting, see \citet{2025arXiv250200111S}.

In general, the cores of merger products are dominated by the initially less massive stars, while the cores of collision products contain some material from the initially more massive stars (see also Section \ref{sec:overview}), and this is reflected in the core hydrogen fractions $\Xc$. The largest differences in core compositions are seen in the 8 + 7, 10 + 7, and 8 + 5 cases, where $\Xc$ of the merger products are higher by $\sim 3 \%$, $\sim 6 \%$, and $\sim 10 \%$, respectively, than the corresponding collision products. The differences in $\Xc$ between collision and merger products in the other cases are within $1 \%$ of each other.

The outermost modeled regions of all collision and merger products have $X \sim 0.70$, which is the hydrogen fraction of isolated MS stellar envelopes. Therefore, we conclude that neither collisions nor mergers will show surface hydrogen depletion or, alternatively, helium enhancement.

Previous studies have shown that post–MS evolution, in particular whether a star evolves to a blue or red supergiant, is highly sensitive to the hydrogen profile just above the convective core \citep[e.g.,][]{2019A&A...625A.132S,2022A&A...662A..56K}. Subtle differences between collision and merger products, as seen in our collision and merger models, may therefore lead to distinct post–MS outcomes, which should be further explored using 1D stellar evolution models capable of following their long-term evolution.

\subsection{Rotation and breakup} \label{sec:rotation}

Collision and merger products rotate differentially and at notable fractions of their breakup velocities owing to the transfer of orbital angular momenta of the initial systems of stars to the spin angular momenta of the resulting products. Figure \ref{fig:profile_angVels} shows the average angular frequencies in the midplanes $\omegamid$ for all collision and merger products at the end of the simulations. All resulting products have steeply decreasing profiles of $\omegamid$, resulting in significant differential rotation. The midplanes are defined as regions around the orbital planes within a vertical distance of $0.5 \Rsun$. Figure \ref{fig:profile_angVel_ratios} shows the ratios of $\omegamid$ to the critical (breakup) angular frequencies $\omegacrit = [G m(s) / s^3]^{0.5}$, where $m(s)$ is the enclosed cylindrical mass within the cylindrical radius $s$, for all collision and merger products at the end of the simulations. The $\omegamid$ values reach maxima of $\sim 0.5$ -- $0.6\,\omegacrit$ within $b = 0.25$ collision products and $\sim 0.8$ -- $0.9\,\omegacrit$ within $b = 0.50$ collision products and merger products.

The large spin angular momenta within merger and collision products have significant implications for their long-term evolution. During thermal contraction of the resulting products, large centrifugal forces can potentially result in substantial mass loss unless they efficiently shed their excess spin angular momenta \citep{2001ApJ...548..323S,2005MNRAS.358..716S}. \citet{2005MNRAS.358..716S} suggested that strong magnetic fields generated during coalescence may lead to loss of angular momentum through magnetic braking by winds \citep{1967ApJ...148..217W} or magnetic disk locking \citep{1978ApJ...223L..83G}. In addition, significant spin-down might occur during thermal contraction owing to internal structural readjustments \citep{2019Natur.574..211S}. Given these various processes that affect stellar spins, the exact long-term spin evolution remains uncertain and requires further study.

\begin{figure*}[ht!]
\centering
\includegraphics[width=\textwidth]{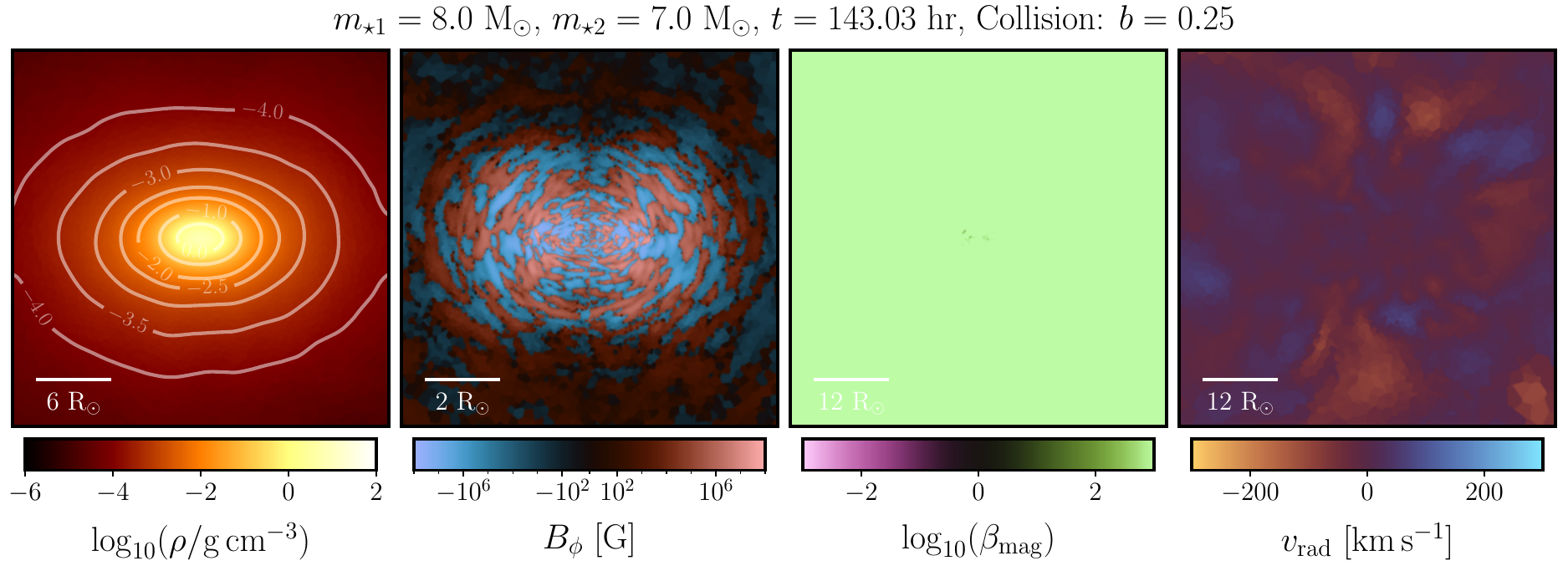}
\includegraphics[width=\textwidth]{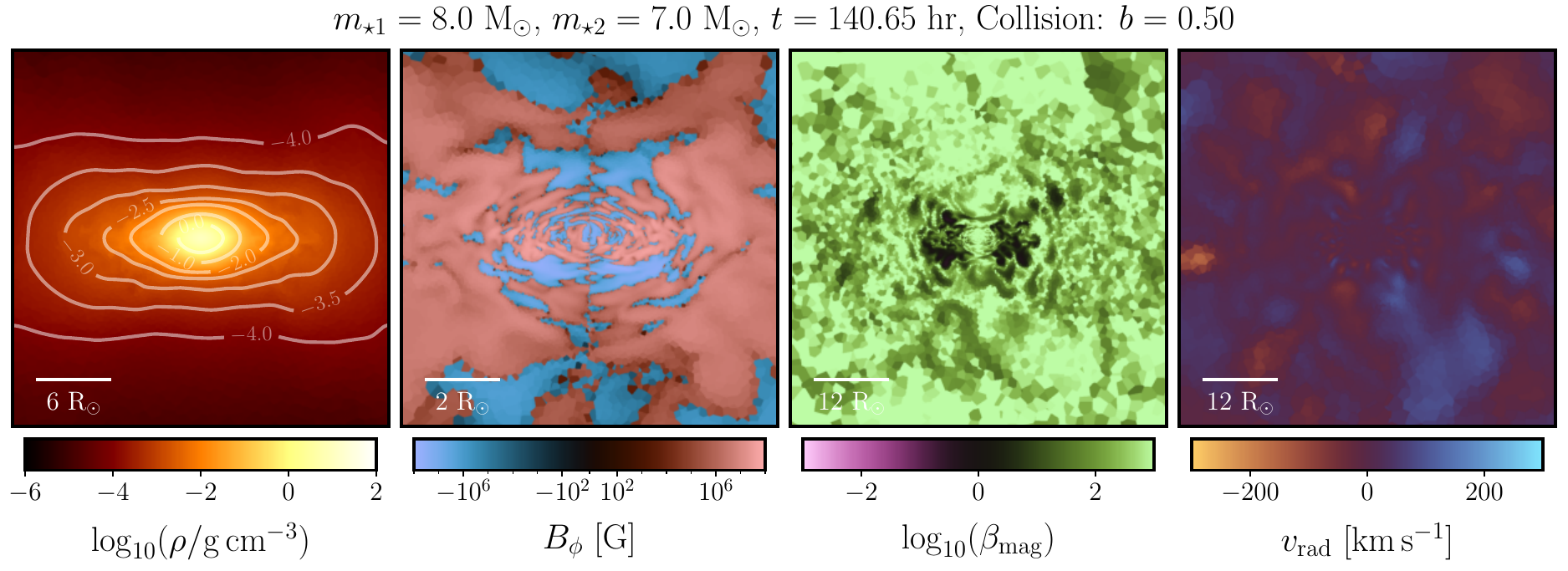}
\includegraphics[width=\textwidth]{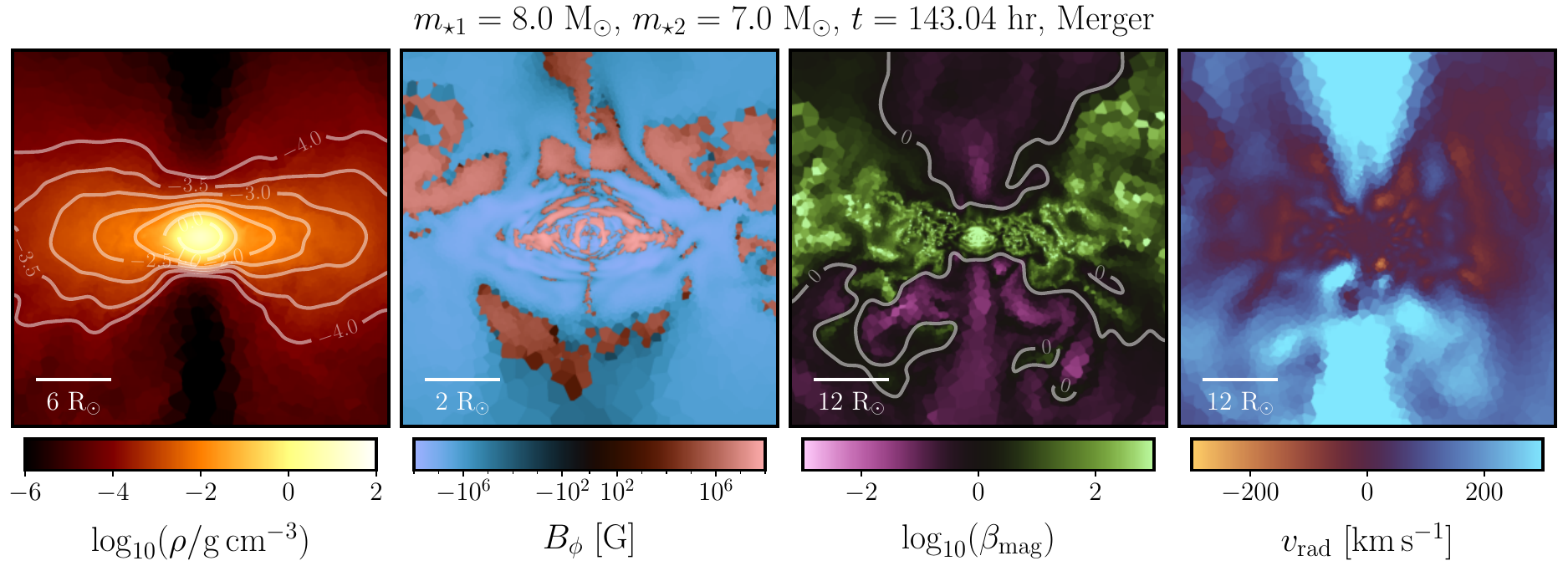}
\caption{Edge-on snapshots of the 8 + 7, $b = 0.25$ collision product (top); 8 + 7, $b = 0.50$ collision product (middle); and the 8 + 7 merger product (bottom) $\sim 60\,\tdynrtot$ after coalescence. All panels illustrate slices, passing through the CoMs, perpendicular to the orbital plane. The four columns illustrate the slices of densities $\rho$ (first column), azimuthal components of magnetic fields $\Bphi$ (second column), plasma beta parameters $\betamag$ (third column), and radial velocities $\vrad$ (fourth column). The color scale of $\Bphi$ is logarithmic when $|\Bphi| \gtrsim 10^2 \G$ and linear otherwise. Also depicted are the smoothed contours of $\rho$ and $\betamag = 1$. The effect of rotational flattening is clearly seen in $\rho$, with the contours of the merger product being peanut shaped. While the $b = 0.25$ collision product shows small-scale reversals in $\Bphi$, the $b = 0.50$ collision product and the merger product have large-scale order. Lastly, magnetized ($\betamag \ll 1$), bipolar axial outflows ($\vrad \gtrsim 300 \kmsinv$) are evident in the merger product but absent in the collision products. \label{fig:coll_mer_collage_fin}}
\end{figure*}

\begin{figure*}[ht!]
\centering
\includegraphics[width=0.48\textwidth]{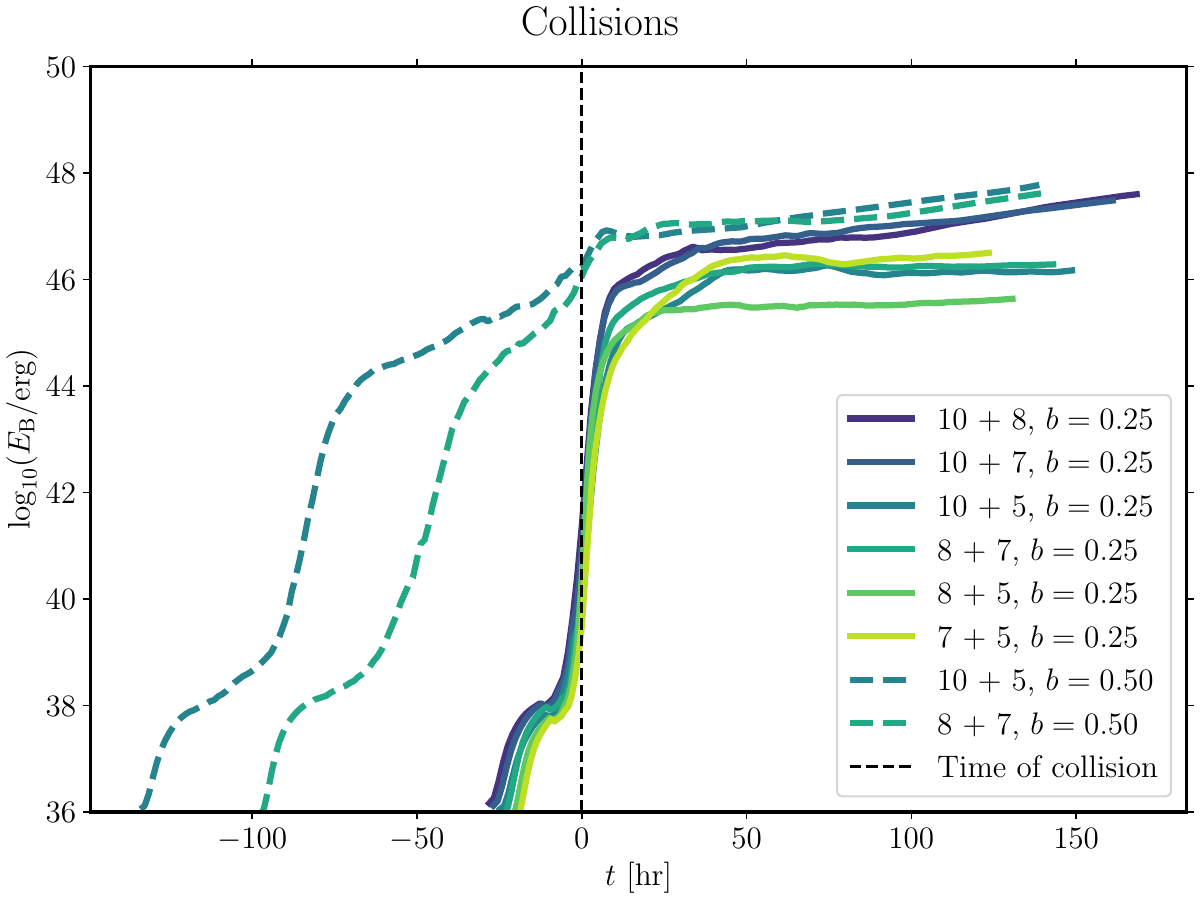}
\includegraphics[width=0.48\textwidth]{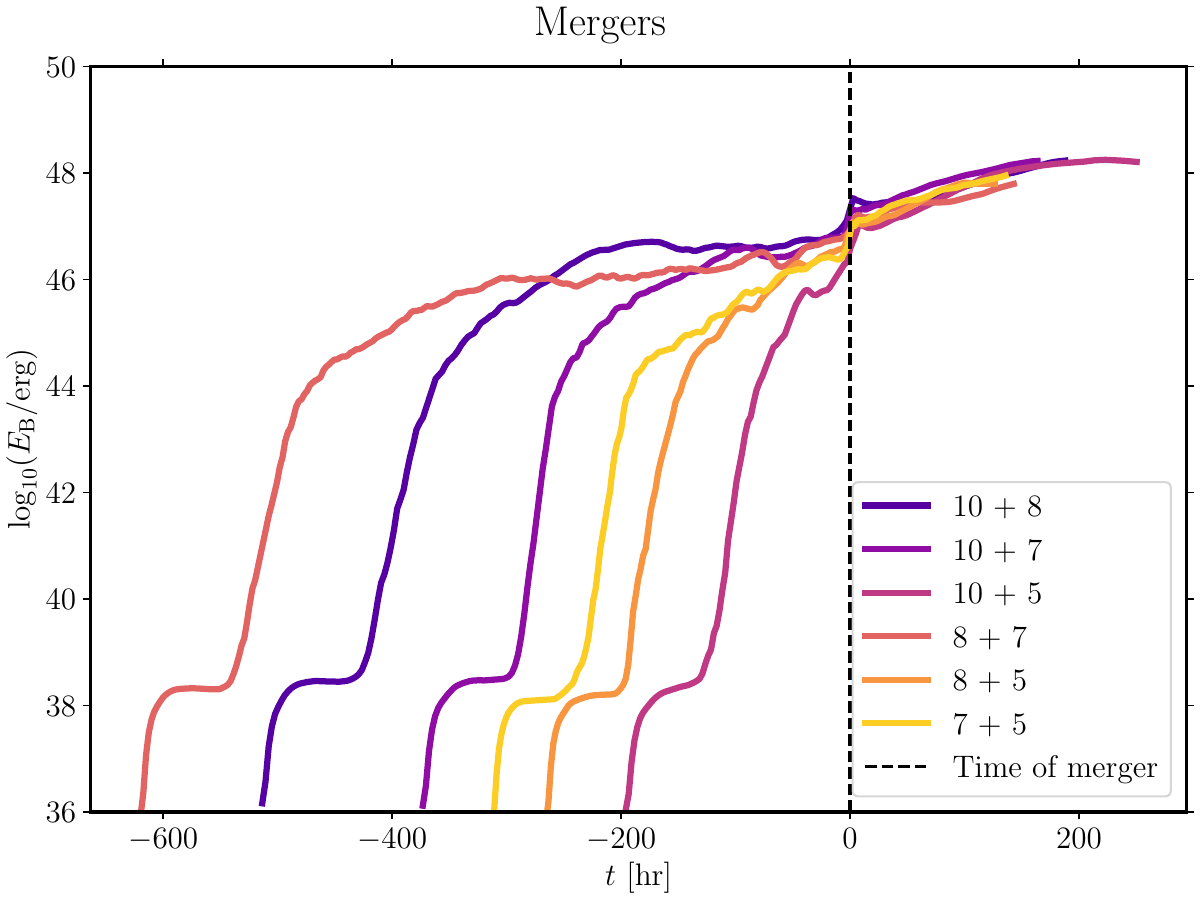}
\caption{Evolution of total magnetic energy $\EB$ through time during collisions (left) and mergers (right) for different masses. For mergers, the times before the actual merger involve artificially accelerated orbit shrinkage. There is an amplification in $\EB$ by $9$ -- $12$ orders of magnitude in both cases. \label{fig:mag_energy_tot}}
\end{figure*}

\subsection{Stellar structures}

Collision and merger products are not spherically symmetric because of rotation, but they are axisymmetric about their rotation axes. Figure \ref{fig:coll_mer_collage_fin} shows the final, edge-on slices of density $\rho$, azimuthal component $\Bphi$ of the magnetic field (detailed in Section \ref{sec:magfield}), plasma beta parameter $\betamag$ (detailed in Section \ref{sec:outflows}), and radial velocity $\vrad$ (detailed in Section \ref{sec:outflows}) for the 8 + 7, $b = 0.25$ collision product (top); the 8 + 7, $b = 0.50$ collision product (middle); and the 8 + 7 merger (bottom). The features of collision and merger products for the other masses are qualitatively similar to those in the 8 + 7 cases (see also Sections \ref{sec:magfield} and \ref{sec:outflows}).

We notice rotational flattening in the edge-on density contours (leftmost column of Figure \ref{fig:coll_mer_collage_fin}) of both collision and merger products. However, while the density contours (and pressure contours; not shown) of the collision products are oblate due to rotation, those of the merger product are peanut shaped, suggesting a dearth of material perpendicular to the merger plane (see also Section \ref{sec:outflows}). We see that the $b = 0.50$ collision product is more flattened \citep[see also][]{2025ApJ...980L..38R} due to a larger initial orbital angular momentum.

\begin{figure*}[ht!]
\centering
\includegraphics[width=0.32\textwidth]{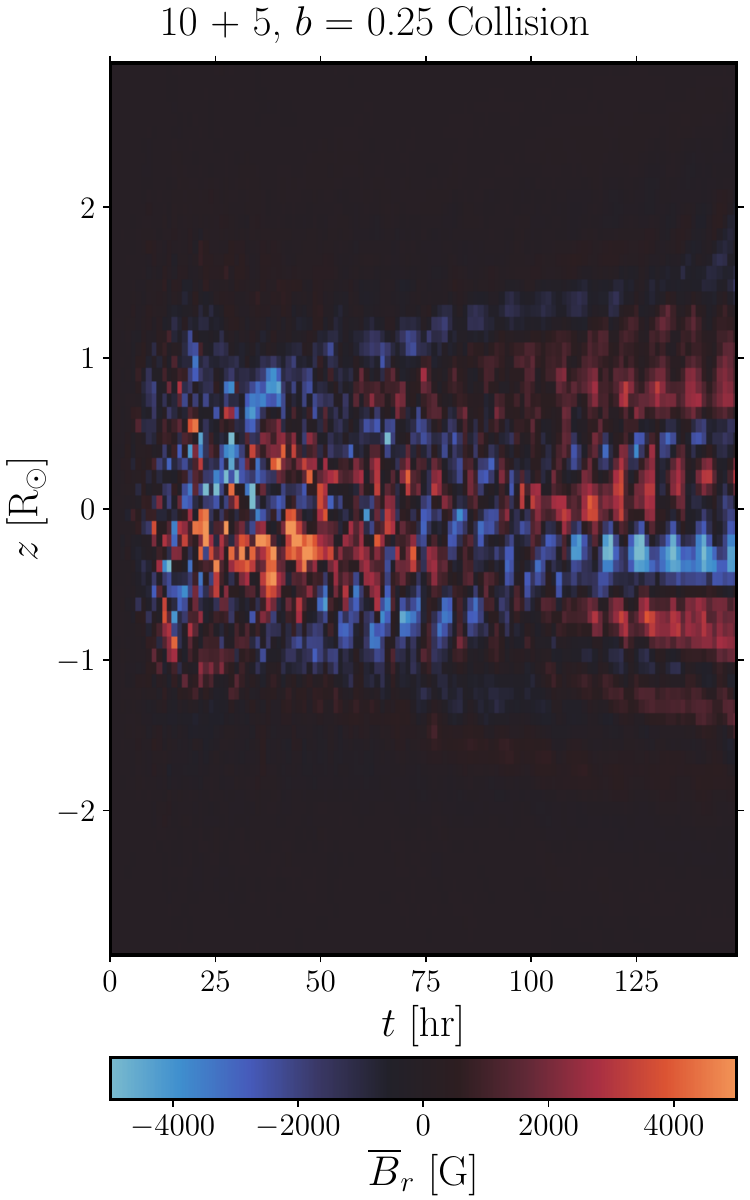}
\includegraphics[width=0.32\textwidth]{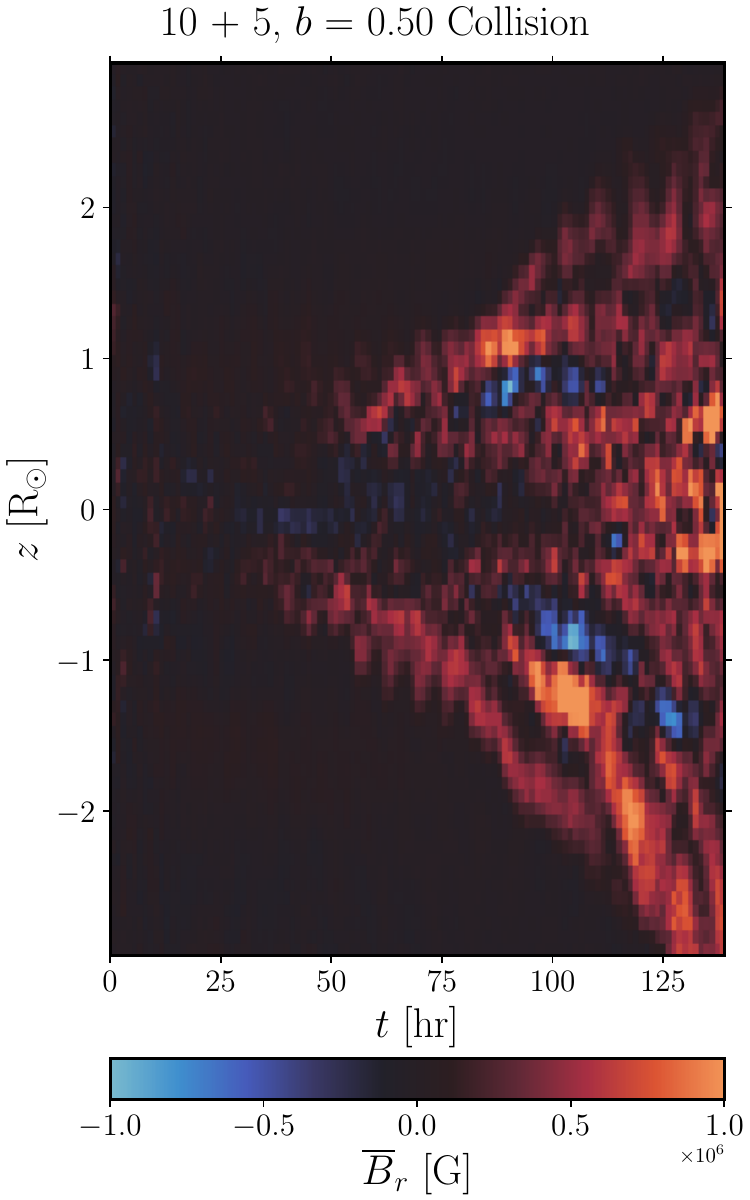}
\includegraphics[width=0.32\textwidth]{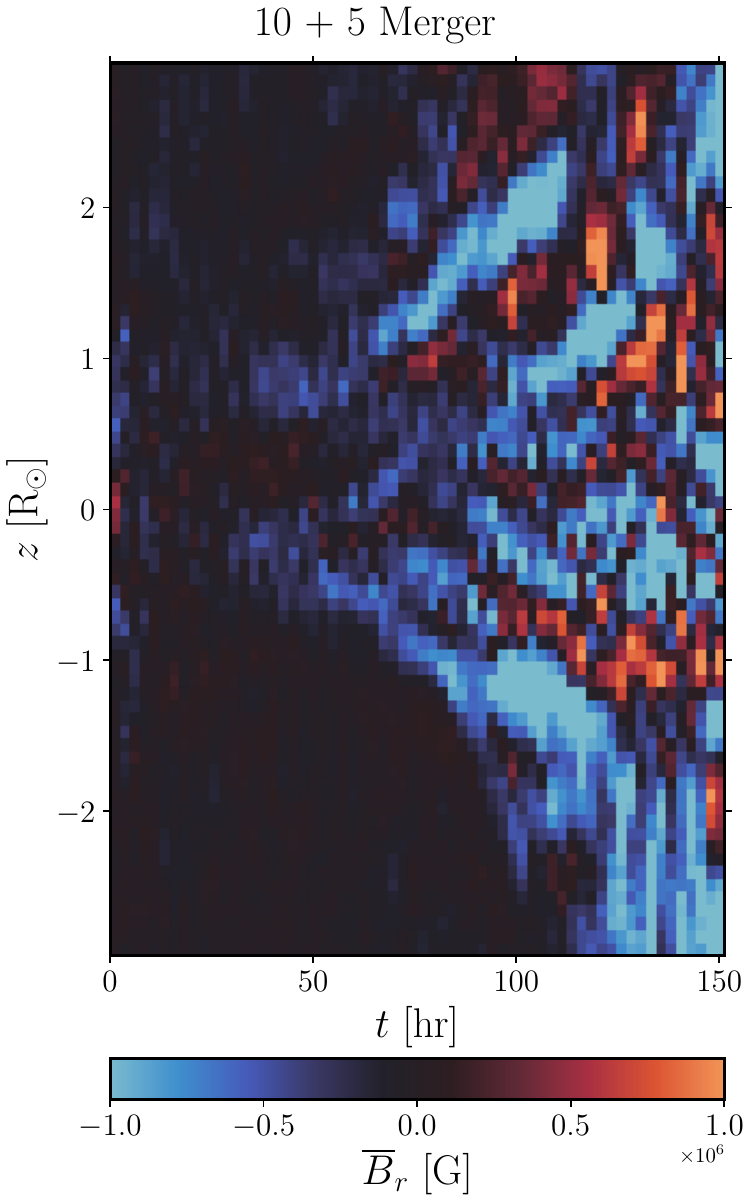}
\caption{Spacetime (`butterfly') diagrams, at cylindrical radii $s \simeq 3 \Rsun$, of the radial components of magnetic field $\Bravg$ (volume averaged). The left, middle, and right panels correspond to the 10 + 5, $b = 0.25$ collision product; the 10 + 5, $b = 0.50$ collision product; and the 10 + 5 merger product respectively. Note that the colorbar scale in the left panel differs from those in the middle and right panels. In the $b = 0.50$ collision product and the merger product, the signs of $\Bravg$ show notable outward-moving patterns, characteristic of MRI. This is not the case in the $b = 0.25$ collision product.} \label{fig:mag_butterfly}
\end{figure*}

\subsection{Magnetic fields} \label{sec:magfield}

In ideal MHD (see Section \ref{sec:method}), magnetic fields are `frozen-in' the stellar plasmas, with no physical magnetic diffusion\footnote{There is, however, numerical diffusion at scales comparable to the sizes of the cells.}. During collisions and mergers, most of the drastic amplification in the magnetic field arises from small-scale dynamos \citep[e.g.,][]{1999ApJ...515L..39C,2016JPlPh..82f5301F,2022A&A...662A..41R}, where turbulent mixing leads to chaotic stretching, folding, and twisting of magnetic field lines. \citet{2019Natur.574..211S} (Extended Data Figure 2) showed that sufficiently resolved simulations of stellar mergers with varying resolutions exhibit similar saturation levels of the total magnetic energy $\EB$, although their (exponential) growth rates vary, lending credence to small-scale dynamos. Additionally, large-scale mixing and the magnetorotational instability (MRI; \citealp{1991ApJ...376..214B}) arising from differential rotation and shear (see Section \ref{sec:rotation}) also lead to some amplification of the magnetic field.

Figure \ref{fig:mag_energy_tot} shows the evolution of $\EB$ as a function of time for collisions and mergers. $\EB$ is amplified by 9 -- 12 orders of magnitude during both types of interaction and saturates at $10^{45}$ -- $10^{48} \erg$. In comparison, the magnitudes of the total energies $|\Etot|$ of the collision and merger products are in the range $(2$ -- $4) \times 10^{50} \erg$ (see Table \ref{tab:new_properties}). In $b = 0.25$ collisions, $\EB$ increases exponentially immediately after the first pericenter passage of the stars and increases sharply again during the second passage and final collision. In the 10 + 8 and 10 + 7 collisions, $\EB$ continues to increase to $\gtrsim 10^{47} \erg$ at the end of the simulations, while it saturates at $\sim 10^{45}$ -- $10^{47} \erg$ in the other collisions. The $b = 0.50$ collisions have larger $\EB$ that also continue to grow owing to turbulent forcing from multiple passages. In the case of mergers, $\EB$ increases during the inspiral phase, saturates momentarily, and increases exponentially again after mass transfer begins. During the merger, $\EB$ surges by an order of magnitude and continues to increase more slowly than before. $\EB$ saturates at $\gtrsim 10^{48} \erg$, as seen for the 10 + 5 merger that we run for an extra $\sim 40\,\tdynrtot$. $\EB$ continues to rise for the other merger simulations, but they all trend toward similar saturation values. Overall, the magnetic field amplification is larger in $b = 0.50$ collisions and mergers than in $b = 0.25$ collisions by $\sim 1$ -- $3$ orders of magnitude.

To understand the source of this magnetic field amplification, we estimate the turbulent energies, $\Eturb$, as the sum of non-azimuthal kinetic energies of the bound stellar material, i.e, $\displaystyle \Eturb \simeq \sum_{\mathrm{cells}} 0.5\,m_{\mathrm{cell}}\,(v_{\mathrm{s,cell}}^2 + v_{\mathrm{z,cell}}^2)$, where $v_{\mathrm{s,cell}}$ and $v_{\mathrm{z,cell}}$ are the cylindrical radial and vertical velocities, respectively, of each cell\footnote{A more accurate turbulent energy includes deviation from rotation in the azimuthal direction. However, this term is not included for simplicity.}. We find that $\EB$ of the 7 + 5, 8 + 5, 8 + 7, and 10 + 5 collisions with $b = 0.25$ is a factor of a few lower than $\Eturb$, suggesting that their magnetic field amplification is primarily sourced from small-scale dynamos that arise due to turbulence (see also \citealp{2025arXiv251213424O,2025arXiv250200111S}). However, in the other collisions and all mergers, $\EB$ is either similar to or a factor of a few higher than $\Eturb$, hinting at other mechanisms, such as MRI, in play.

We test this hypothesis of MRI augmenting magnetic field amplification in Figure \ref{fig:mag_butterfly}, which presents the space-time (or `butterfly') diagrams \citep[e.g.,][]{1995ApJ...446..741B,1996ApJ...463..656S} of the radial components of magnetic field $\Bravg$ for the 10 + 5 collision and merger products. Such plots are quite useful for diagnosing the differences in magnetic fields (and other quantities) at different vertical heights $z$ over time $t$. To generate space-time diagrams, we first consider cells in a cylindrical region of radius $s \simeq 3 \Rsun$ around the axis perpendicular to the orbital plane and passing through the CoM, for each simulation snapshot after coalescence. Next, we bin these selected cells based on their heights relative to the orbital plane. Finally, we volume-average the radial components of magnetic fields in each bin and plot this quantity, $\Bravg$, as a function of $z$ and $t$.

Figure \ref{fig:mag_butterfly} shows that $\Bravg$ changes sign and rises above the midplane over time in the $b = 0.50$ collision product and the merger product but not in the $b = 0.25$ collision product. These vertically propagating fields may be the result of (large-scale) dynamo waves \citep{1955ApJ...122..293P}, similar to the $11 \yr$ Solar sunspot cycle \citep[e.g.,][]{1961ApJ...133..572B}, which are driven by differential rotation and MRI. In such dynamos, MRI and turbulence generate poloidal fields ($\alpha$ effect), and differential rotation generates toroidal fields ($\Omega$ effect), both of which are necessary for stable magnetic field configurations \citep{1973MNRAS.163...77M,1973MNRAS.161..365T,2004Natur.431..819B}. Interestingly, unlike other $b = 0.25$ collision products, this signature of MRI also develops in the 10 + 8 and 10 + 7 collision products $\sim 100 \hr$ after collision, supporting the larger (and gradually increasing) $\EB$ in these two cases.

Lastly, we explore the longevity of magnetic fields by examining the field morphology. A magnetic field with many small-scale reversals can diminish in strength over time owing to magnetic reconnection. In contrast, a field with large-scale order, in addition to differential rotation, may be powered by a large-scale dynamo and can indicate longevity \citep[e.g.,][]{2024NatAs...8..298K,2024A&A...691A.179P,2025arXiv250918421S}. The second column of Figure \ref{fig:coll_mer_collage_fin} shows the edge-on slices of the toroidal component of the magnetic field, $\Bphi$, for the 8 + 7 collision and merger products at the end of the simulations. Whereas the $b = 0.25$ collision product shows many small-scale reversals in $\Bphi$, the $b = 0.50$ collision product and the merger product do not. Small-scale reversals are also observed in all $b = 0.25$ collision products, and large-scale order is seen in $b = 0.50$ collision products (especially in the outer regions) and merger products of different masses. Crucially, this implies that grazing (large impact parameter) collisions, akin to mergers, result in stars that can potentially generate large-scale dynamos and sustain magnetic fields.

\begin{figure*}[ht!]
\centering
\includegraphics[width=0.48\textwidth]{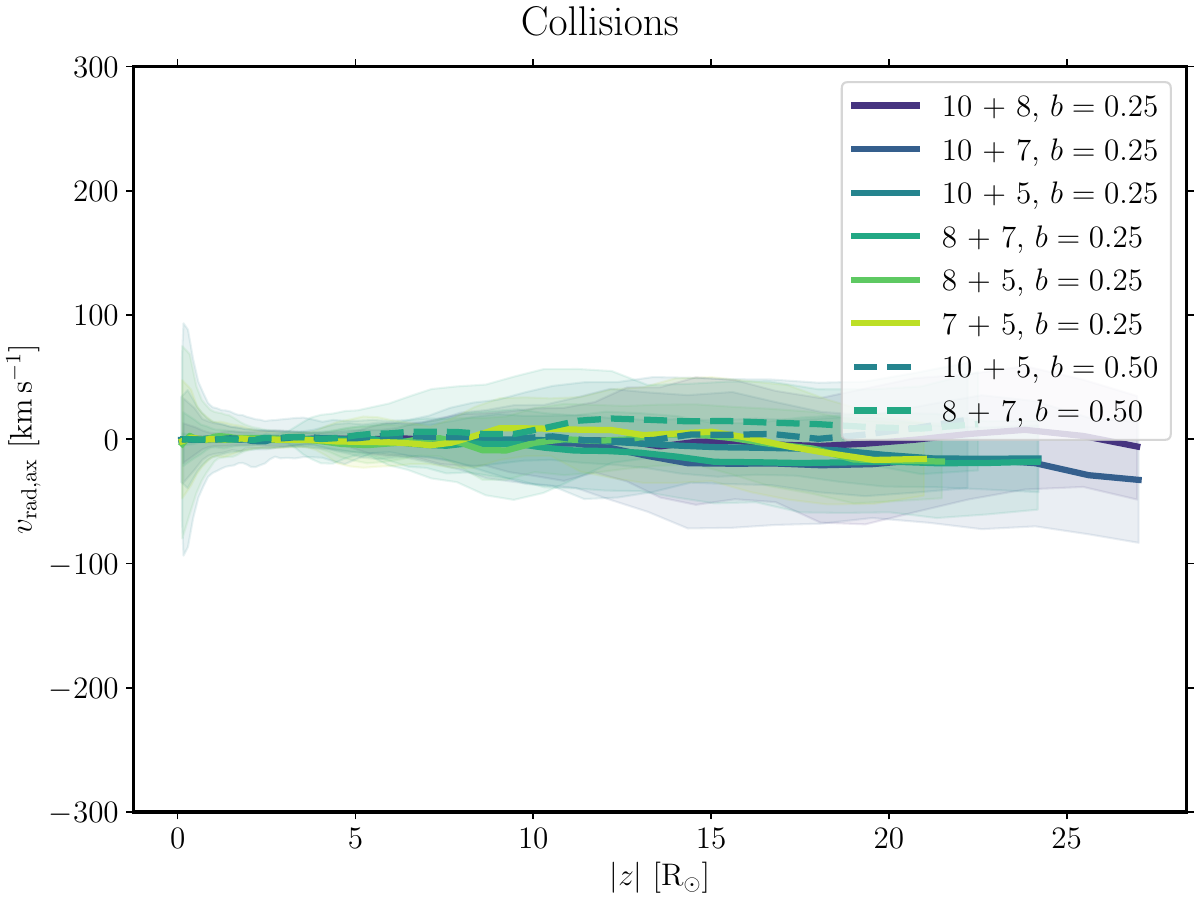}
\includegraphics[width=0.48\textwidth]{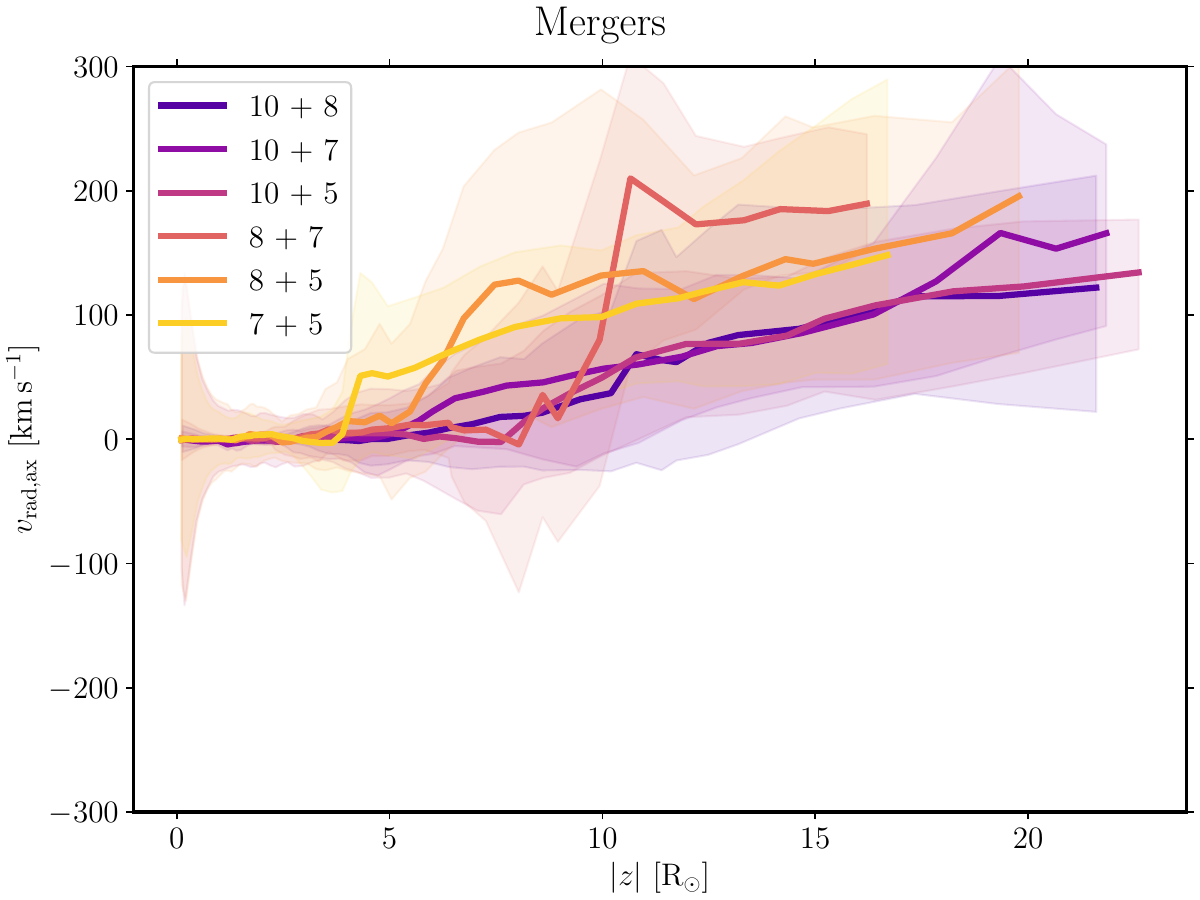}
\caption{Profiles of average radial velocities in the bipolar axial regions (see text for definition) $\vradax$ as functions of vertical (axial) distances $|z|$. The left and right panels represent collision and merger products, respectively, of different masses. Merger products have clear axial outflows with velocities $\vradax \sim 100$ -- $300 \kmsinv$ while collision products do not.} \label{fig:profile_radVels}
\end{figure*}

\begin{deluxetable*}{cccc}[ht!]
\tablecaption{Summary of the differences between collision products of different impact parameters and merger products. \label{tab:summary_tab}}
\tablehead{
\colhead{Properties} & \colhead{Collisions: $b = 0.25$ } & \colhead{Collisions: $b = 0.50$} & \colhead{Mergers}
}
\startdata
Maximal angular frequency & $\sim (0.5$ -- $0.6)\,\omegacrit$ & $\sim (0.8$ -- $0.9)\,\omegacrit$ & $\sim (0.8$ -- $0.9)\,\omegacrit$ \\
Shapes of density contours & Oblate & More flattened oblate  & Peanut shaped \\
Magnetic field energies & $\sim 10^{45}$ -- $10^{48} \erg$ & $\gtrsim 10^{48}\erg$ & $\gtrsim 10^{48}\erg$ \\
Magnetic field configurations & Small-scale reversals & Large-scale ordered & Large-scale ordered \\
Magnetized outflows & No & No & Yes ($\betamag \sim 10^{-1}$ -- $10^{-2}$) \\
\enddata
\tablecomments{Here, $b$, $\omegacrit$, and $\betamag$ are the collision impact parameter, critical angular frequency, and the plasma beta parameter, respectively.}
\end{deluxetable*}

\subsection{Magnetized bipolar outflows} \label{sec:outflows}

A quantitative indicator of the significance of magnetic fields in gas dynamics is the plasma beta parameter, defined as $\betamag = P/P_{\mathrm{mag}}$. Here, $P$ is the thermal pressure (gas + radiation) and $P_{\mathrm{mag}} = |\bm{B}|^{2} / 8\pi$ is the magnetic pressure. Thus, $\betamag \ll 1$ suggests that magnetic pressure dominates gas dynamics and vice versa. The third column of Figure \ref{fig:coll_mer_collage_fin} shows the magnetic $\betamag$ of the surrounding material for the 8 + 7 collision and merger products at the end of the simulations. $\betamag > 10^4$ ($\gg 1$) for the $b = 0.25$ collision product, indicating that its magnetic field plays no role in gas dynamics. $\betamag \sim 10$ -- $100$ for the $b = 0.50$ collision product, notably higher and implying that its magnetic field plays a minor role in gas dynamics. However, the nonaxial and axial regions of the merger products have $\betamag \sim 1$ -- $100$ and $\sim 0.01$ -- $1$, respectively, underscoring the importance of magnetic fields in governing gas dynamics, most notably in driving bipolar axial outflows.

Figure \ref{fig:profile_radVels} shows the average radial velocities in the bipolar axial regions for all collision and merger products at the end of the simulations. The axial regions are defined as bicones of $30^{\circ}$ centered on the CoMs of the remnants and oriented perpendicular to the orbital planes. We see a stark disparity in their radial velocities: the collision products do not show strong, directed outflows ($\vradax < 50 \kmsinv$), while the merger product shows distinct bipolar axial outflows with $\vradax \sim 100$ -- $300 \kmsinv$. This is also evident from the fourth column of Figure \ref{fig:coll_mer_collage_fin}, which shows the edge-on slices of radial velocities $\vrad$ (in all directions) of the surrounding material for the 8 + 7 collision and merger products. In addition, the outflow velocities are higher farther away from the merger products, large enough to escape them and become unbound; the magnetized outflows carry away $\sim 5 \%$ -- $10 \%$ of the total magnetic energies of the merger products.

Many MHD simulations of stellar interactions \citep[e.g.,][]{2022A&A...660L...8O,2024NatAs...8..298K,2024A&A...681A..41M,2024A&A...691A.179P,2025A&A...698A.133V} have shown magnetically driven bipolar outflows of plasma. Although the exact origin of outflows is unclear, they require a strong magnetic field at the center and a long-lived interaction to develop and collimate gas along the rotation axis. We speculate that the difference in the bipolar outflows between collision and merger products may partly be explained by the timescale of interaction. A collision occurs in a few impulsive stellar passages, ranging from $\sim 10 \hr$ when $b = 0.25$ to $\sim 80$ -- $100 \hr$ when $b = 0.50$, which does not seem to be long enough for magnetized outflows to develop and collimate. In contrast, stars undergoing a binary merger interact for much longer timescales (see Figure \ref{fig:mag_energy_tot}), ranging from $\sim 200 \hr$ for the 10 + 5 merger to $\sim 600 \hr$ for the 8 + 7 merger\footnote{Given a constant inspiral velocity, stars with a more skewed mass ratio merge more quickly than those with a near-equal mass ratio.}, during the accelerated inspiral phase. In reality, their timescales of interaction would be much longer than the orbital timescales, which would provide sufficient time for outflows to develop and strengthen. However, a more detailed investigation of the physical origin of outflows is beyond the purview of this study.

\section{Summary and conclusions} \label{sec:conclude}

The aim of this work is to compare the differences and similarities between two types of stellar interactions: collisions and mergers of two MS stars. Employing the 3D moving-mesh code \arepo{}, we conducted 14 high-resolution MHD simulations of nonrotating MS stars realistically modeled using the 1D stellar evolution code \mesa{}: 8 collisions and 6 mergers for stellar masses ranging from $5 \Msun$ to $10 \Msun$. In the case of collisions, we set the impact parameters $b = \rp/\rtot$ of the initial parabolic orbits at $b = 0.25$ (closer to head-on) for 6 simulations and at $b = 0.50$ (farther from head-on) for 2 simulations. In the case of mergers, we assumed initially circular binary orbits that finally merge owing to artificially accelerated inspiral. Our main results are detailed below, while Table \ref{tab:summary_tab} provides a concise summary of the major differences between collision and merger products.

\begin{itemize}
    \item A direct consequence of stellar collisions and mergers is the chemical mixing of stars, resulting in hydrogen in the initial stellar envelopes being mixed into the new stellar cores and helium in the initial stellar cores being mixed into the new stellar envelopes. Although the cores are rejuvenated in both types of interactions, the exact core hydrogen fractions in the merger products can be $3 \%$ -- $10 \%$ higher than the corresponding collision products in some cases. Thus, their MS lifetimes are also longer by a similar fraction, which may affect future stellar evolution. 
    \item Collision and merger products are fast and differentially rotating owing to the conversion of orbital angular momenta into spin angular momenta. The $b = 0.50$ collision products and merger products rotate up to $0.8$ -- $0.9$ of breakup velocity, while $b = 0.25$ collision products rotate slower at $0.5$ -- $0.6$ of breakup velocity.
    \item As a result of rotation, collision and merger products are flattened. The degree of rotational flattening in the collision products is higher for larger impact parameter ($b = 0.50$) encounters. In comparison, the density contours of merger products are peanut shaped, with a dearth of material in the polar regions.
    \item During the course of stellar interactions, turbulent mixing results in exponential amplifications of magnetic field energies $\EB$ by $9$ -- $12$ orders of magnitude assuming a small initial field. The $b = 0.25$ collision products have systematically lower $\EB$, by $1$ -- $3$ orders of magnitude, than the $b = 0.50$ collision products and merger products.
    \item The azimuthal components of magnetic fields, $\Bphi$, show small-scale reversals for $b = 0.25$ collision products, implying that the magnetic fields may be relatively short-lived. On the contrary, $b = 0.50$ collision products and merger products show large-scale order in $\Bphi$, necessary to generate large-scale dynamos and sustain long-lasting magnetic fields.
    \item Magnetized, bipolar axial outflows are  launched only from merger products, driven by strong magnetic fields, whereas we find no sign of outflows in collisions. The plasma beta parameters of these outflows are $\betamag \sim 10^{-1}$ -- $10^{-2}$, with outflow velocities of up to $\vrad \sim 300 \kmsinv$.
\end{itemize}

In future work, we plan to model the stellar evolution of collision and merger products in detail to understand the differences in their long-term evolution.

\begin{acknowledgments}
P.V. thanks Robert Mathieu and Selma de Mink for helpful discussions. P.V. thanks Bart Ripperda and Norman Murray for providing access to their computational time on Compute Canada's \textit{Niagara} cluster. Most of the simulations presented in this paper were carried out using computational resources (and/or scientific computing services) at the Max-Planck Computing \& Data Facility.
\end{acknowledgments}

\begin{contribution}
PV performed the simulations in this paper and led the analysis and writing. All authors contributed to the project planning, interpretation of the analysis, and writing.
\end{contribution}

\facilities{DRAC: \textit{Niagara} cluster}

\software{\arepo{} \citep{2010MNRAS.401..791S,2016MNRAS.455.1134P,2020ApJS..248...32W}, \mesa{} \citep{2011ApJS..192....3P}}



\bibliography{blue_strag}{}
\bibliographystyle{aasjournalv7}

\end{document}